\documentclass{article}
\usepackage[utf8]{inputenc}
\usepackage{macros}

\usepackage{graphicx}
\graphicspath{ {./images/} }

\newcommand{\OPT}{\mathsf{OPT}}

\newcommand{\OL}{\mathsf{OL}}

\title{Root Ridge Leverage Score Sampling for $\ell_p$ Subspace Approximation}
\author{
David P. Woodruff \\ Carnegie Mellon University \\ \texttt{dwoodruf@cs.cmu.edu} \and 
Taisuke Yasuda \\ The Voleon Group\thanks{Part of this work was done while T.Y. was at Carnegie Mellon University.} \\ \texttt{yasuda.taisuke1@gmail.com}
}

\begin{document}

\maketitle

\thispagestyle{empty}
\begin{abstract}

The $\ell_p$ subspace approximation problem is an NP-hard low rank approximation problem that generalizes the median hyperplane problem ($p = 1$), principal component analysis ($p = 2$), and the center hyperplane problem ($p = \infty$). A popular approach to cope with the NP-hardness of this problem is to compute a \emph{strong coreset}, which is a small weighted subset of the input points which simultaneously approximates the cost of every $k$-dimensional subspace, typically to $(1+\varepsilon)$ relative error for a small constant $\varepsilon$.

We obtain an algorithm for constructing a strong coreset for $\ell_p$ subspace approximation of size $\tilde O(k\varepsilon^{-4/p})$ for $p<2$ and $\tilde O(k^{p/2}\varepsilon^{-p})$ for $p>2$. This offers the following improvements over prior work:
\begin{itemize}
    \item We construct the first strong coresets with nearly optimal dependence on $k$ for all $p\neq 2$. In prior work, \cite{SW2018} constructed coresets of modified points with a similar dependence on $k$, while \cite{HV2020} constructed true coresets with polynomially worse dependence on $k$. 
    \item We recover or improve the best known $\varepsilon$ dependence for all $p$. In particular, for $p > 2$, the \cite{SW2018} coreset of modified points had a dependence of $\varepsilon^{-p^2/2}$ and the \cite{HV2020} coreset had a dependence of $\varepsilon^{-3p}$.
\end{itemize}
Our algorithm is based on sampling by root ridge leverage scores, which admits fast algorithms, especially for sparse or structured matrices. Our analysis completely avoids the use of the representative subspace theorem \cite{SW2018}, which is a critical component of all prior dimension-independent coresets for $\ell_p$ subspace approximation.

Our techniques also lead to the first nearly optimal \emph{online} strong coresets for $\ell_p$ subspace approximation with similar bounds as the offline setting, resolving a problem of \cite{WY2023b}. All prior approaches lose $\mathrm{poly}(k)$ factors in this setting, even when allowed to modify the original points.

\end{abstract}

\clearpage
\setcounter{page}{1}

\section{Introduction}

\subsection{The \texorpdfstring{$\ell_p$}{lp} subspace approximation problem}

Consider a set of $n$ points $\{\bfa_1, \bfa_2, \dots, \bfa_n\}$ in $d$-dimensional space. A natural problem is to determine the ``cost'' of moving these $n$ points to lie on a low-dimensional subspace $F\subseteq\mathbb R^d$ of rank $k$. This is an old and important question that has been studied independently in varied fields such as operations research, statistics, and computational geometry. %

We measure the cost of moving a single point to a subspace $F$ by the Euclidean distance, given by
\[
    \cost(\bfa_i, F) = \min_{\bfx\in F}\norm{\bfa_i - \bfx}_2 = \norm{\bfa_i^\top(\bfI - \bfP_F)}_2
\]
where $\bfI$ denotes the $d\times d$ identity and $\bfP_F$ denotes the orthogonal projection matrix onto $F$. This cost is rotationally invariant, which is important for accurately modeling real world data \cite{DZHZ2006}. To measure the cost of the entire set of $n$ points, we aggregate the cost of the individual points via the \emph{$\ell_p$ loss}, which is parameterized by some exponent $p\geq 1$ and is given by
\[
    \cost^p(\{\bfa_i\}_{i=1}^n, F) = \bracks*{\sum_{i=1}^n \norm{\bfa_i^\top(\bfI-\bfP_F)}_2^p}^{1/p}.
\]
For smaller values of $p$, this cost tends to capture the average cost of a point $\bfa_i$, while for larger values of $p$, this cost tends to capture the maximum cost of a point $\bfa_i$. The problem of minimizing this objective over rank $k$ subspaces $F$ is known as the \emph{$\ell_p$ subspace approximation problem}. To simplify the notation, we will write the objective as simply $\norm{\bfA(\bfI-\bfP_F)}_{p,2}$, where $\bfA$ is the $n\times d$ matrix containing $\bfa_i$ in the $i$-th row and $\norm{\cdot}_{p,2}$ is the \emph{$(p,2)$-norm}, which is defined for a $n\times d$ matrix $\bfY$ as $\norm{\bfY}_{p,2} \coloneqq \bracks{\sum_{i=1}^n \norm{\bfe_i^\top\bfY}_2^p}^{1/p}$. We will denote the optimal value of this problem as
\[
    \OPT \coloneqq \min_{F\in\mathcal F_k} \norm{\bfA(\bfI-\bfP_F)}_{p,2}^p
\]
where $\mathcal F_k$ denotes the set of all rank $k$ subspaces $F$. We will let $\bfP^*$ denote the projection matrix for a rank $k$ subspace achieving this optimum.

Several special cases of this problem are particularly well-studied. The most popular special case is the sum of squared costs, or the \emph{$\ell_2$ loss}, which is the usual \emph{principal component analysis (PCA) problem} and is one of the most popular dimension reduction techniques in statistics and data analysis. In modern machine learning and statistical practice, the $\ell_1$ subspace approximation problem has been studied intensely as a natural variant of the PCA that retains rotational invariance yet offers solutions which are robust to outliers \cite{SW1987, DZHZ2006, GZAL2014, YHY2017, NNSS2020}. In transportation theory, the $\ell_1$ subspace approximation cost for $k = 1$ is studied as the problem of solving for the cheapest way to lay out a railroad (i.e., a 1-dimensional subspace) to serve $n$ cities \cite{MB1972, MN1980}. In computational geometry, the $\ell_1$ subspace approximation problem is also known as the \emph{median hyperplane problem} \cite{MS2001}, and also has a rich history \cite{KM2012, Sch1999, MS2001}. A closely related problem is the regression problem, for which \cite{sun2024introduction} motivates the study of $\ell_p$ loss for $p$ strictly between $1$ and $2$, i.e., $1 < p < 2$, due to a combination of its robustness and efficiency properties.

In the other extreme, the $\ell_\infty$ subspace approximation problem is particularly well-studied in the computational geometry literature, where it is known as the \emph{center hyperplane problem} \cite{MS2001}, and generalizes a variety of \emph{containment} or \emph{extent} problems such as minimum enclosing balls and minimum enclosing cylinders \cite{GK1994, HV2004, AHV2004}. However, the $\ell_\infty$ norm poses various computational challenges \cite{DTV2011, GRSW2012} and may be too sensitive to extreme values, and the $\ell_p$ norm for large values of $p$ are often considered as useful relaxations of the $\ell_\infty$ norm problem that behave more smoothly \cite{SV2012}. For related regression problems, values of $p$ strictly between $2$ and $\infty$, i.e., $2 < p < \infty$, offer a compromise between robustness and accuracy, and find applications in the design of
polynomial finite impulse response filters  \cite{burrus1994iterative, dumitrescu2007positive}.

The flexibility of allowing for arbitrary values of $p$ is also crucial for tuning the behavior of the optimization problem to a specific input distribution. As a toy example, consider an input matrix $\bfA = \bfB + \bfG$, where $\bfB$ is a $n\times n$ diagonal matrix with $k$ ones and $n-k$ zeros on the diagonal, and $\bfG$ is i.i.d.\ Gaussian noise with variance $\sigma^2$. Then, the $(p,2)$-norm of $\bfG$ to the power $p$ is roughly $n^{p/2+1}\sigma^p$ while the $(p,2)$-norm of $\bfA$ to the power $p$ is $k((1+\sigma)^2 + (n-1) \sigma^2)^{p/2} + (n-k) n^{p/2} \sigma^p$, that is, larger by roughly an additive $k$. Thus, in order for the $\ell_p$ subspace approximation problem to output a good approximation to $\bfB$, $p$ must be tuned such that $k = \Omega(n^{p/2+1}\sigma^p)$; otherwise, the $0$ matrix is already a constant factor approximation. We stress that {\it any setting of $p$} could be important here, depending on the variance $\sigma^2$ of $\bfG$. For instance, if $\sigma^2 = k^{2/3}/n^{5/3}$, then we should set $p = 3$, and if $\sigma^2 = k^{1/2}/n^{3/2}$, then we should set $p = 4$. Note that we also do not want to set $p$ larger than we have to, as we shall see that the complexity of the problem increases for large $p$. 

\subsection{Sparsification and coresets}

In big data settings, the number of points $n$ is often prohibitively large.  \emph{Sparsification} and \emph{coresets} are a class of techniques which have been extremely successful in handling such parameter regimes. Sparsification generally refers to the approximation of a large object by a sparse object, and has been studied extensively in settings such as graphs \cite{BK1996, ST2004, BSS2012} and hypergraphs \cite{BST2019, KKTY2021, JLS2023, Lee2023} as well as a large variety of decomposable loss functions such as clustering \cite{Che2009, FL2011, BFL2016, FSS2020, HV2020, BJKW2021, CSS2021, CLSS2022, CLSSS2022, HLW2022}, submodular functions \cite{RY2022, JLLS2023}, $\ell_p$ regression \cite{DMM2006, Cla2005, DDHKM2009, CP2015, WY2023a}, and generalized linear models \cite{MMR2021, MOP2022, MMWY2022, JLLS2024}. When the object being sparsified is a dataset, then its sparse approximation is often called a coreset. Coresets can often be used as a substantially smaller drop-in replacement of the original dataset, and leads to both time- and space-efficient algorithms.

Coresets have been a particularly important tool in the study of algorithms for $\ell_p$ subspace approximation, due to the fact that the problem is known to be NP-hard for any $p\neq 2$ \cite{DTV2011, GRSW2012, CW2015a}. In fact, the best known $(1+\eps)$ approximation algorithms for this problem proceed exactly by reducing the input size via coreset techniques, and then applying an exponential time algorithm on the reduced instance \cite{FMSW2010, CW2015a}. Thus, the best known running time of $\ell_p$ subspace approximation is almost completely determined by how small the coreset can be made to be. Various coreset algorithms for $\ell_p$ subspace approximation have been developed for both $p = 2$ \cite{DMM2006b, DMM2008, CEMMP2015, CMM2017} and $p\neq 2$ with increasingly improved sizes, running times, and guarantees \cite{DV2007, FL2011, SV2012, VX2012, SW2018, FSS2020, HV2020, FKW2021}, generalized loss functions \cite{CW2015a, MMWY2022}, and in information-limited settings such as streaming \cite{FMSW2010, LSW2018, MRWZ2020, DP2022} and online \cite{BLVZ2019, BDMMUWZ2020, WY2023b} models.

\subsubsection{Strong coresets for \texorpdfstring{$\ell_p$}{lp} subspace approximation}
\label{sec:intro:strong-coresets}

While several different coreset guarantees have been studied, we focus on \emph{strong coresets}, which are a weighted subset of points $\bfa_i$ such that the weighted cost of any rank $k$ subspace $F$ on the coreset approximates the cost on the original dataset, up to a factor of $(1+\eps)$.

\begin{Definition}[Strong coresets for $\ell_p$ subspace approximation]
\label{def:strong-coreset}
Let $1\leq p < \infty$ and $0 < \eps < 1$. Let $\bfA\in\mathbb R^{n\times d}$. Then, a diagonal matrix $\bfS\in\mathbb R^{n\times n}$ is a \emph{$(1\pm\eps)$ strong coreset} for $\ell_p$ subspace approximation if
\[
    \norm*{\bfS\bfA(\bfI-\bfP_F)}_{p,2}^p = (1\pm\eps)\norm*{\bfA(\bfI-\bfP_F)}_{p,2}^p
\]
for every $F\in\mathcal F_k$. We refer to the number of nonzero entries $\nnz(\bfS)$ of $\bfS$ as the size of the coreset.
\end{Definition}

The guarantee of Definition \ref{def:strong-coreset} can also be viewed as a natural generalization of \emph{projection cost-preserving sketches} \cite{CEMMP2015, CMM2017, MM2020} to the $\ell_p$ subspace approximation setting. 

Strong coresets can be used to reduce the size of the input instance to at most $\nnz(\bfS)$ points in $\nnz(\bfS)$ dimensions by rotational invariance. In particular, strong coresets of size $\nnz(\bfS) = \poly(k/\eps)$ immediately remove the dependence of this problem on $n$ and $d$, making this a powerful tool in the design of fast algorithms.

It is useful to compare the guarantee of Definition \ref{def:strong-coreset} with other popular coreset guarantees. One weaker guarantee is a \emph{weak coreset}, which requires that if $\tilde F$ is the optimal solution to the subspace approximation problem for $\bfS\bfA$, then it is also a $(1+\eps)$-optimal solution for $\bfA$ \cite{FL2011, HV2020}. Another further weaker guarantee is a \emph{spanning coreset}, which requires that there is a rank $k$ subspace $\tilde F$ in the row span of $\bfS\bfA$ that is a $(1+\eps)$-optimal solution \cite{DV2007, SV2012, CW2015a, WY2024}. The guarantee of Definition \ref{def:strong-coreset} immediately achieves both of these guarantees, and offers further benefits that cannot be realized by these other guarantees, such as applications to solving \emph{constrained} versions of $\ell_p$ subspace approximation, or solving $\ell_p$ subspace approximation in distributed and streaming models via the merge-and-reduce technique \cite{BDMMUWZ2020, CWZ2023}. Known algorithms for a number of constrained low rank approximation problems such as binary matrix factorization \cite{VVWZ2023}, sparse PCA \cite{Pia2023}, projective non-negative matrix factorization \cite{YO2005}, and constrained subspace estimation \cite{SVKMPS2017} all start by first computing a coreset, as a coreset preserves all possible solutions, and then use a problem-dependent algorithm on the coreset whose running time is proportional to the coreset size. By reducing the coreset size for $\ell_p$-subspace approximation, we can improve the time complexity for all of these constrained problems. 

\subsubsection{Prior work}

While it has long been known that strong coresets of size $\poly(k,d,\eps^{-1})$ independent of $n$ exist \cite{FL2011}, the first dimension-independent result, i.e., a strong coreset of size only $\poly(k,\eps^{-1})$ independent of $d$, was achieved by the work of \cite{SW2018}. This result achieves a coreset of size of $\nnz(\bfS) = \tilde O(k)\eps^{-O(1)}$ for $p < 2$ and $\nnz(\bfS) = \tilde O(k^{p/2})\eps^{-O(p^2)}$ for $p>2$, which is in fact nearly optimal with respect to $k$ \cite{LWW2021}. However, this result has several drawbacks. First, the $\eps$ dependence is poor, with an $O(p^2)$ dependence in the exponent which makes this result intractable for even moderately large $p$. Second, it does not quite satisfy Definition \ref{def:strong-coreset}, due to the fact that the coreset is a weighted subset of points \emph{with an appended coordinate} which must be treated separately from the original coordinates, rather than a weighted subset of the original data points themselves. This is unfavorable for a number of reasons:
\begin{itemize}
    \item The use of an appended coordinate deviates from the traditional definition of coresets, which is the standard for a vast majority of the coreset literature.
    \item Coreset size lower bounds against row subsets no longer apply to these constructions, which are often substantially easier to establish than general data structure lower bounds. For instance, there are several coreset lower bounds which are specific to row subsets such as \cite{HV2020, CLSS2022, HLW2022} for clustering. Another example is \cite{LWW2021}, which obtains nearly optimal (row subset) coreset lower bounds for $\ell_p$ subspace embeddings, and looser lower bounds for general data structures.
    \item Downstream algorithms for $\ell_p$ subspace approximation must be modified to handle the additional coordinate, which may not always be simple. Algorithms for subspace approximation are highly varied, ranging from convex relaxation and rounding \cite{DTV2011} to power iterations \cite{DZHZ2006, NNSS2020} to polynomial solvers \cite{CW2015a}, and it is undesirable to adapt all of these analyses and implementations.
    \item In the online coreset model, in which the coreset must be constructed ``on the fly'' as the input points $\bfa_i$ arrive, this approach leads to suboptimal bounds \cite{WY2023b} (see Section \ref{sec:intro:online}).
\end{itemize}
Furthermore, the algorithm of \cite{SW2018} runs in exponential time. The running time was reduced to polynomial time in a follow-up work of \cite{FKW2021}, but this result loses $\poly(k)$ factors in the coreset size and still has the $\eps^{-O(p^2)}$ dependence and the extra appended coordinate.

Among purely row subset coresets following Definition \ref{def:strong-coreset}, the work of \cite{HV2020} gives the first dimension-independent strong coresets using a technique known as sensitivity sampling, which can be implemented in input sparsity time \cite{FKW2021, WY2023b}. However, the coreset size in this work, while dimension-independent, is not optimal, and loses $\poly(k)$ factors in the coreset size $\nnz(\bfS)$. Thus, there is a $\poly(k, \eps^{-1})$ factor gap in our understanding of the complexity of strong coresets for $\ell_p$ subspace approximation, and it is a central question in this literature to pin down the dependence of $k$. In particular, it is natural to conjecture that the \cite{SW2018} result can be matched with a purely row subset coreset.

\begin{Question}
\label{q:optimal-coreset}
Do strong coresets for $\ell_p$ subspace approximation of size $\tilde O(k)\poly(\eps^{-1})$ for $p<2$ and $\tilde O(k^{p/2})\poly(\eps^{-1})$ for $p>2$ exist? Can such coresets be constructed efficiently?
\end{Question}

\subsection{Our contributions}

The main result of this work is a positive resolution to Question \ref{q:optimal-coreset} for all $1\leq p < \infty$. For $1\leq p < 2$, we obtain the first row subset strong coreset with a nearly linear dependence on $k$. In particular, for $p = 1$, we obtain strong coresets of size $\tilde O(k/\eps^4)$ for the median hyperplane problem, which is off by only a $1/\eps^2$ factor off from the lower bound of $\tilde\Omega(k/\eps^2)$ \cite{LWW2021, WY2023b}, and improves the previous upper bound of $\tilde O(k^4 / \eps^6)$ due to \cite{HV2020}. For $p = 2$, the work of \cite{CMM2017} already gives a bound of $\tilde O(k/\eps^2)$ for ridge leverage score sampling. However, we offer a new proof of this fact that is arguably much simpler.

\begin{Theorem}%
\label{thm:strong-coreset-p<2}
Let $1\leq p<2$. Let $\bfA\in\mathbb R^{n\times d}$. Then, there is an algorithm running in $\tilde O\parens*{\nnz(\bfA) + d^\omega}$ time which, with probability at least $1-\delta$, constructs a strong coreset $\bfS$ of size
\[
    \nnz(\bfS) = \frac{k}{\eps^{4/p}}(\log(k/\eps\delta))^{O(1)}
\]
satisfying Definition \ref{def:strong-coreset}, that is,
\[
    \norm*{\bfS\bfA(\bfI-\bfP_F)}_{p,2}^p = (1\pm\eps)\norm*{\bfA(\bfI-\bfP_F)}_{p,2}^p \qquad \mbox{for every subspace $F\subseteq\mathbb R^d$ of rank at most $k$}.
\]
\end{Theorem}

For $2 < p < \infty$, we obtain a bound that scales as $k^{p/2}$. While this is not nearly linear, this dependence on $k$ is necessary for this range of $p$, and matches a lower bound of $\tilde\Omega(k^{p/2}\eps^{-1} + k\eps^{-2})$ \cite{LWW2021, WY2023b}.

\begin{Theorem}%
\label{thm:strong-coreset-p>2}
Let $2 < p < \infty$. Let $\bfA\in\mathbb R^{n\times d}$. Then, there is an algorithm running in $\tilde O\parens*{\nnz(\bfA) + d^\omega}$ time which, with probability at least $1-\delta$, constructs a strong coreset $\bfS$ of size
\[
    \nnz(\bfS) = \frac{k^{p/2}}{\eps^{p}}(\log(k/\eps\delta))^{O(p^2)}
\]
satisfying Definition \ref{def:strong-coreset}, that is,
\[
    \norm*{\bfS\bfA(\bfI-\bfP_F)}_{p,2}^p = (1\pm\eps)\norm*{\bfA(\bfI-\bfP_F)}_{p,2}^p \qquad \mbox{for every subspace $F\subseteq\mathbb R^d$ of rank at most $k$}.
\]
\end{Theorem}

We give several remarks concerning our results. First, we are the first to establish even the \emph{existence} of a weighted subset of points with nearly optimal dependence on $k$, and we are able to construct such a subset in nearly input sparsity time. We also note that for $p<2$, the fact that we achieve nearly linear size for the strong coreset guarantee implies that we simultaneously achieve the first nearly optimal size guarantees for other weaker yet popular coreset guarantees, including weak coresets \cite{FL2011, HV2020}. That is, for $p<2$, we are in fact the first to resolve Question \ref{q:optimal-coreset} even for weaker notions of coresets, both for existence and efficient constructions. For $p>2$, we obtain the best known construction for weak coresets and the best known efficient construction for spanning coresets. Finally, as we see in more depth later, our algorithm is based off of sampling rows by using scores known as \emph{ridge leverage scores}, which is a primitive known to be have highly efficient algorithms, especially when the underlying matrix has additional structure such as efficient matrix-vector products \cite{SW2019} or positive semidefiniteness \cite{MW2017}.

\begin{table}[ht]
\label{tab:prior}
\caption{Coreset size upper bounds for $\ell_p$ subspace approximation. Some bounds in prior work have been sharpened using our Appendix \ref{sec:sharper-sohler-woodruff}.}
\centering
\begin{tabular}{ l l l l l }
\toprule
 & Coreset size & Coreset guarantee & Extra coordinate & Running time \\
\midrule
\cite{FL2011} & $k^{\max\{1,p/2\}}\eps^{-2}\cdot k^3\eps^{-1}$ & weak & -- & input sparsity \\
\cite{FL2011} & $k^{\max\{1,p/2\}}\eps^{-2}\cdot dk$ & strong & -- & input sparsity \\
\cite{SW2018} & $k^{\max\{1,p/2\}}\eps^{-\max\{4,p^2/2+2\}}$ & strong & yes & exponential \\
\cite{FKW2021}\tablefootnote{In Appendix \ref{sec:fast-sw}, we show that the ideas of this work can in fact be improved to input sparsity running time and coreset size $k^{\max\{1,p/2\}}\eps^{-\Theta(p^2)}$. However, the extra coordinate and significantly worse dependence on $\epsilon$ remain.} & $k^3\eps^{-6}$, $p = 1$ & strong & yes & polynomial \\
\cite{HV2020} & $k^{\max\{1,p/2\}+3}\eps^{-\max\{6,3p\}}$ & strong & -- & input sparsity \\
\rowcolor{blue!15} This work & $k^{\max\{1,p/2\}}\eps^{-\max\{4/p,p\}}$ & strong & -- & input sparsity \\
\bottomrule
\end{tabular}
\end{table}

\subsubsection{Nearly optimal online coresets for \texorpdfstring{$\ell_p$}{lp} subspace approximation}
\label{sec:intro:online}

Despite the several drawbacks discussed in Section \ref{sec:intro:strong-coresets}, strong coresets that use an additional coordinate as in \cite{SW2018, FKW2021} may still be acceptable for applications to time- and space-efficient algorithms. This is especially true since we have shown that these approaches can be implemented in input sparsity time with nearly optimal dependence on $k$ on the coreset size in Appendix \ref{sec:fast-sw}. However, we discuss one application of our results which requires our new approach to this problem, which is an application to sampling strong coresets for $\ell_p$ subspace approximation in the \emph{online coreset model}.

The literature of online algorithms studies algorithms which must make irrevocable decisions as its input arrives incrementally. In online machine learning and data analysis, a typical setting is for new data points to arrive one by one. The algorithm must then immediately make a decision about this data point, such as assigning them to clusters \cite{Mey2001} or outputting a prediction \cite{Haz2016}. The online coreset model, introduced by work of \cite{CMP2016}, studies the question of implementing coreset algorithms in such online settings. More specifically, the rows of our design matrix $\bfA\in\mathbb R^{n\times d}$ arrive one by one in a stream, and we must sample the rows of the coreset in an online fashion, that is, we must decide whether or not to keep a row $i\in[n]$ when it arrives, and we cannot discard rows that we keep, or retrieve rows that we discard. This model was originally studied in the context of coresets for $\ell_2$ linear regression \cite{CMP2016}, and has subsequently been analyzed for a variety of problems in data analysis, computational geometry, and numerical linear algebra \cite{BLVZ2019, BDMMUWZ2020, WY2022b, WY2023a, WY2023b, MMO2024}.

For $\ell_p$ subspace approximation in particular, the works of \cite{BLVZ2019, BDMMUWZ2020} studied the case of $p = 2$ and achieved a nearly optimal coreset size of $\tilde O(k\eps^{-2})$, while \cite{WY2023b} studied the case of $p\neq 2$, achieving a coreset size of roughly $\tilde O(k^{p + O(1)})\eps^{-\Theta(p)}$ by analyzing an algorithm based on sensitivity sampling \cite{HV2020}.\footnote{In our discussion of online coresets, we allow for the $\tilde O(\cdot)$ notation to suppress polylogarithmic factors in $n$ and an ``online condition number'' quantity $\kappa^\OL$ which appears in all prior works on online coresets and is known to be necessary.} One of the main open questions left in \cite{WY2023b} is to obtain a nearly optimal dependence on $k$ for online strong coresets for $\ell_p$ subspace approximation.

\begin{Question}
\label{q:online-coreset}
Are there online coreset algorithms for constructing strong coresets for $\ell_p$ subspace approximation of size $\tilde O(k)\poly(\eps^{-1})$ for $p<2$ and $\tilde O(k^{p/2})\poly(\eps^{-1})$ for $p>2$?
\end{Question}

In the online coreset setting, all prior techniques for dimension-free coresets \cite{SW2018, HV2020, FKW2021} face a key challenge. The problem is that all of these techniques are based on computing a fixed low-dimensional subspace $S$ which is used to construct the coreset. In the online coreset setting, this is a major obstruction, due to the fact that this low-dimensional subspace $S$ can evolve as more inputs arrive. In the recent work of \cite{WY2023b}, it was shown that this problem can be circumvented by bounding the number of times that a subspace can undergo a significant change, using the so-called \emph{online Lewis weights} \cite{WY2023a}. However, this technique increases the coreset size by the number of times this subspace changes, which is a factor of $\tilde O(k^{\max\{1,p/2\}})$, leading to a suboptimal overall bound of $\tilde O(k^{p + O(1)})\poly(\eps^{-1})$. In particular, the extra factor of $k^{p/2}$ is a glaring gap in our understanding of this problem, as pointed out by \cite{WY2023b}.

Despite these challenges, our work gives a positive resolution of Question \ref{q:online-coreset} by introducing a completely new approach towards coresets for $\ell_p$ subspace approximation. One of the key aspects of our techniques will be to avoid a ``commitment'' to a specific low dimensional subspace via the use of regularization, and will allow us to seamlessly handle the online coreset setting. The technical details are given in Section \ref{sec:online-coresets}.

\begin{table}[ht]
\label{tab:prior-online}
\caption{Online/streaming coreset size upper bounds for $\ell_p$ subspace approximation.}
\centering
\begin{tabular}{ l l l l l }
\toprule
 & Coreset size & Coreset guarantee & Extra coordinate & Streaming model \\
\midrule
\cite{BLVZ2019} & $k\eps^{-2}$, $p = 2$ & spanning & -- & online coreset \\
\cite{BDMMUWZ2020} & $k\eps^{-2}$, $p = 2$ & strong & -- & online coreset \\
\cite{DP2022} & $k\cdot (k\eps^{-1})^{p+1}$ & spanning & -- & row arrival \\
\cite{WY2023b} & $k^{p + O(1)}\eps^{-\Theta(p)}$ & strong & -- & online coreset \\
\rowcolor{blue!15} This work & $k^{\max\{1,p/2\}}\eps^{-\max\{4/p,p\}}$ & strong & -- & online coreset \\
\bottomrule
\end{tabular}
\end{table}

\subsection{Technical overview}

We will now give a technical overview of our results. In the following discussion, $\bfS$ will always be constructed as a \emph{$\ell_p$ sampling matrix}, with varying choices of sampling probabilities $q_i$.

\begin{Definition}[$\ell_p$ sampling matrix]
\label{def:sampling-matrix}
Let $p\geq 1$. A random diagonal matrix $\bfS\in\mathbb R^{n\times n}$ is an \emph{$\ell_p$ sampling matrix with sampling probabilities $\{q_i\}_{i=1}^n$} if for each $i\in[n]$, the $i$th diagonal entry is independently set to be
\[
    \bfS_{i,i} = \begin{cases}
        1 / q_i^{1/p} & \text{with probability $q_i$} \\
        0 & \text{otherwise}
    \end{cases}
\]
\end{Definition}

\subsubsection{Pitfalls in prior work}
\label{sec:pitfalls}

The central technique of \cite{SW2018} is a structural result which shows the existence of a \emph{representative subspace} $S\subseteq\mathbb R^d$ with $s = O(k)\poly(\eps^{-1})$ dimensions such that for any $k$-dimensional subspace $F\subseteq\mathbb R^d$,
\[
    \norm*{\bfA(\bfI-\bfP_F)}_{p,2}^p = (1\pm\eps)\norm*{[\bfA\bfP_S(\bfI-\bfP_F), \bfb_S]}_{p,2}^p
\]
where $\bfb_S\in\mathbb R^n$ is the vector with $i$-th entry given by $\bfb_S(i) = \norm*{\bfa_i^\top(\bfI-\bfP_S)}_2$, and $[\bfA\bfP_S(\bfI-\bfP_F), \bfb_S]$ is the $n\times (d+1)$ matrix formed by the concatenation of $\bfA\bfP_S(\bfI-\bfP_F)$ and $\bfb_S$.\footnote{We improve the analysis of this result by removing a $1/\eps^3$ factor from $s$ in Appendix \ref{sec:sharper-sohler-woodruff}.} That is, the $\ell_p$ subspace approximation cost of $F$ can be approximated by the projection cost onto the subspace $S$, plus the additional projection cost of the lower dimensional points $\bfA\bfP_S$ to the query subspace $F$. This reduces the subspace approximation problem in $d$ dimensions to a similar problem in $s+1$ dimensions. In turn, this lower dimensional problem can be solved using dimension-dependent techniques, since the dimension is now only $s+1 = O(k)\poly(\eps^{-1})$. Then to analyze sampling algorithms, \cite{SW2018} show that Dvoretzky's theorem \cite{Dvo1961, FLM1977, PVZ2017} can be applied to convert the problem of approximating the $(p,2)$-norm to a problem of approximating the $(p,p)$-norm, i.e. the entrywise $\ell_p$ norm, which can then be handled by sampling techniques for approximating $\ell_p$ norms of vectors in a subspace \cite{CP2015, WY2023a}. This shows that
\[
    \norm*{\bfS[\bfA\bfP_S(\bfI-\bfP_F), \bfb_S]}_{p,2}^p = (1\pm\eps)\norm*{[\bfA\bfP_S(\bfI-\bfP_F), \bfb_S]}_{p,2}^p
\]
when $\bfS$ is constructed as a subspace embedding for the column space of $[\bfA\bfP_S, \bfb_S]$. Ideally, we would like to apply the same reasoning as before to argue that
\[
    \norm*{\bfS\bfA(\bfI-\bfP_F)}_{p,2}^p = (1\pm\eps)\norm*{\bfS[\bfA\bfP_S(\bfI-\bfP_F), \bfb_S]}_{p,2}^p
\]
to complete the chain of approximations. Unfortunately, it is unclear whether this step goes through when $\bfS$ is constructed as a subspace embedding of $[\bfA\bfP_S, \bfb_S]$. The intuition for this is that $S$ is constructed to be a representative subspace for $\bfA$ using information about $\bfA$ \emph{outside} of the column space of $[\bfA\bfP_S, \bfb_S]$, and there is no guarantee that a sampling matrix $\bfS$ that preserves $[\bfA\bfP_S, \bfb_S]$ also preserves this information.

On the other hand, the result of \cite{HV2020} takes a different approach based on the classic \emph{sensitivity sampling} technique \cite{LS2010, FL2011, VX2012}, and uses the representative subspace constructed \cite{SW2018} in an \emph{existential} manner rather than an algorithmic one. For sensitivity sampling, one first defines \emph{sensitivity scores} 
\begin{equation}\label{eq:sensitivities}
    \bfsigma_i(\bfA) \coloneqq \sup_{F\in\mathcal F_k}\frac{\norm*{\bfa_i^\top(\bfI-\bfP_F)}_2^p}{\norm*{\bfA(\bfI-\bfP_F)}_{p,2}^p}
\end{equation}
for each row $i\in[n]$ which represent the largest fraction of the cost occupied by the $i$-th coordinate, ranging over all queries $F\in\mathcal F_k$. Then, by Bernstein bounds, it follows that for any fixed $F\in\mathcal F_k$, sampling the rows $i\in[n]$ proportionally to the sensitivity scores preserves $\norm*{\bfA(\bfI-\bfP_F)}_{p,2}^p$ up to $(1\pm\eps)$ factors. We can then apply this result to every $F$ in a net over the space of rank $k$ subspaces $F$, which has size roughly $\exp(dk)$, to construct coresets of size $\poly(d,k,\eps^{-1})$. The work of \cite{HV2020} improves this argument by showing that the representative subspace $S$ constructed by \cite{SW2018} allows one to derive a strong coreset guarantee from a weak coreset guarantee, which in turn admit dimension-independent bounds \cite{FL2011}. However, the weak coreset argument of \cite{FL2011} loses $\poly(k)$ factors.

\subsubsection{Ridge leverage scores}

Our algorithmic technique takes a drastically different approach compared to the prior works of \cite{SW2018, HV2020, FKW2021}. Our starting point is a result of \cite{CMM2017}, which resolves Question \ref{q:optimal-coreset} for the much simpler case of $p = 2$. For $p = 2$, finding an explicit rank $O(k)\poly(\eps^{-1})$ subspace with properties similar to the representative subspace $S$ is not difficult due to the singular value decomposition (SVD) \cite{DMM2006b, DMM2008, CEMMP2015, CMM2017}. However, as noted by \cite{CMM2017}, while this gives a polynomial time algorithm for low rank approximation for $p = 2$, finding these scores is already as hard as low rank approximation itself. Thus, this defeats the purpose of finding the coreset if the goal is to design faster algorithms. To address this problem, \cite{CMM2017} make use of the following alternative scores for a sampling-based algorithm, known as the \emph{ridge leverage scores}.

\begin{Definition}[Ridge leverage scores \cite{AM2015, CMM2017}]
Let $\lambda>0$ and $\bfA\in\mathbb R^{n\times d}$. Then, for each $i\in[n]$, the $i$th ridge leverage score is defined as
\[
    \bftau_i^\lambda(\bfA) \coloneqq \bfa_i^\top(\bfA^\top\bfA + \lambda\bfI)^{-1}\bfa_i = \sup_{\bfx\in\mathbb R^d} \frac{[\bfA\bfx](i)^2}{\norm{\bfA\bfx}_2^2 + \lambda\norm{\bfx}_2^2}.
\]
\end{Definition}

Note that ridge leverage scores can be approximated up to $O(1)$ factors in just $\tilde O(\nnz(\bfA) + d^\omega)$ time, where $\omega$ is the exponent of matrix multiplication \cite{SS2011, DMMW2012, CW2013, CLMMPS2015}.

The main result of \cite{CMM2017} establishes that if we set $\lambda = \norm*{\bfA-\bfA_k}_F^2/k$ where $\bfA_k$ is the best rank $k$ approximation in the Frobenius norm, then sampling $\tilde O(k/\eps^2)$ rows $\bfa_i$ of $\bfA$ proportionally to their ridge leverage scores yields a strong coreset $\bfS$ of nearly optimal size satisfying Definition \ref{def:strong-coreset} for $p = 2$. Furthermore, the scores $\bftau_i^\lambda(\bfA)$ only depend on a constant factor approximation to the \emph{value} of the optimal low rank approximation, which can be obtained more readily than a subspace which (approximately) witnesses this value. Unfortunately, the analysis of \cite{CMM2017} is highly specific to the $\ell_2$ norm, and makes heavy use of the structural properties of the SVD and the fact that the $(p,2)$-norm is an entrywise norm for $p = 2$, and thus does not apply to $p\neq 2$. Nonetheless, several key ideas still do carry over to the setting of $p\neq 2$, which will be crucial to our analysis. 

\subsubsection{Reduction to embedding low rank matrices}

We start with a conceptual simplification of the problem. Our goal is to preserve the objective function $\norm{\bfA(\bfI-\bfP_F)}_{p,2}^p$ for all rank $k$ subspaces $F$. We will write this as $\norm{\bfA(\bfP^* - \bfP_F) + \bfA(\bfI-\bfP^*)}_{p,2}^p$ where $\bfP^*$ is the optimal rank $k$ projection. That is, we think the objective as the sum of a ``low rank'' component and a ``small fixed'' component. Due to the bound on $\norm{\bfA(\bfI-\bfP^*)}_{p,2}^p$ and the fact that it is fixed, adding the ``small fixed'' component turns out to be relatively straightforward; technically, it amounts to replacing guarantees for embedding subspaces to embedding affine spaces. As shown in prior work \cite{MMWY2022, WY2024}, this can be done even when the affine component is unknown, as long as the sampling probabilities are large enough. We thus focus on preserving the ``low rank'' component $\norm{\bfA(\bfP^* - \bfP_F)}_{p,2}^p$ via sampling, that is, we focus on the guarantee that
\[
    \norm{\bfS\bfA(\bfP^* - \bfP_F)}_{p,2}^p = (1\pm\eps)\norm{\bfA(\bfP^* - \bfP_F)}_{p,2}^p.
\]
In fact, we will only need to use the fact that $\bfP^* - \bfP_F$ has low rank and bounded operator norm. Thus, we generalize this to the guarantee that
\[
    \norm{\bfS\bfA\bfX}_{p,2}^p = (1\pm\eps)\norm{\bfA\bfX}_{p,2}^p
\]
for every $\bfX$ with rank at most $2k$ and operator norm at most $2$. Finally, a crucial observation is that our original objective function is always at least $\OPT$, so we can afford an \emph{additive} error of $\eps\OPT$. That is, it suffices to show the guarantee that
\begin{equation}\label{eq:generalized-guarantee}
    \norm{\bfS\bfA\bfX}_{p,2}^p = (1\pm\eps)\norm{\bfA\bfX}_{p,2}^p \pm \eps \OPT
\end{equation}
in our setting. Thus for the rest of this technical overview, we will focus on showing \eqref{eq:generalized-guarantee}.

Note that at this point, our approach already differs from the prior dimension-independent coreset results of \cite{SW2018} and \cite{HV2020}, in that we \emph{directly} tackle the problem of approximating the original objective function via sampling, rather than approximating the objective by a lower-dimensional objective via a representative subspace. This is a critical step: rather than relying on the representative subspace to sample small coresets, we instead rely on the allowance of a small additive error of $\eps \OPT$ in sampling primitives, which in turn allow for smaller sample sizes along with other benefits.

\begin{Remark}
In an earlier version of this work, the guarantee of \eqref{eq:generalized-guarantee} was used in a different way which ensured that (1) the coreset was a subspace embedding for a fixed subspace $S$ of dimension $k/\eps^{\max\{2,p\}}$, and (2) the coreset guaranteed that $\norm{\bfS\bfA(\bfP_{S\cup F} - \bfP_S)}_{p,2}^p = \norm{\bfA(\bfP_{S\cup F} - \bfP_S)}_{p,2}^p \pm \eps\OPT$ for every rank-$k$ subspace $F$. However, requiring the guarantee of (1) causes the $\eps$ dependence to be much worse, and in particular by plugging in the dimension of $k/\eps^{\max\{2,p\}}$ into known subspace embedding dimension requirements \cite{CP2015, WY2023a}, the coreset would need to be of a much larger size of at least $k^{p/2}/\eps^{p^2/2}$ for $p>2$.
\end{Remark}

\subsubsection{Idea 1: additive-multiplicative \texorpdfstring{$\ell_p$}{lp} subspace embeddings via root ridge leverage scores}
\label{sec:intro:add-mult-se}

We begin by applying Dvoretzky's theorem, which gives a linear map $\bfH$ such that $\norm{\bfH^\top\bfx}_p^p = (1\pm\eps)\norm{\bfx}_2^p$ for every $\bfx\in\mathbb R^d$, so that $\norm{\bfA\bfX}_{p,2}^p = (1\pm \eps)\norm{\bfA\bfX\bfH}_{p,p}^p$. This decouples the norm of the columns, reducing our problem to preserving the $\ell_p$ norm of vectors of the form $\bfA\bfx$, i.e., we wish to show $\norm{\bfS\bfA\bfx}_p^p \approx \norm{\bfA\bfx}_p^p$. If the approximation guarantee is purely multiplicative, i.e. $\norm{\bfS\bfA\bfx}_p^p = (1\pm\eps) \norm{\bfA\bfx}_p^p$, then this guarantee is known in the literature as an \emph{$\ell_p$ subspace embedding} \cite{CP2015, WY2023a}. However, this guarantee is too strong, and requires coreset size depending polynomially on $d$. We thus need a relaxation of this guarantee to achieve small coresets.

The first new ingredient in our analysis is to adapt the \emph{additive-multiplicative $\ell_2$ subspace embedding} idea of \cite{CMM2017}. In this result, \cite{CMM2017} show that a sampling matrix $\bfS$ with probabilities proportional to the ridge leverage scores $\bftau_i^\lambda(\bfA)$ with $\lambda = \norm{\bfA-\bfA_k}_F^2/k$ satisfies the additive-multiplicative guarantee
\[
    \norm*{\bfS\bfA\bfx}_2^2 = (1\pm\eps)\norm*{\bfA\bfx}_2^2 \pm \eps\lambda\norm{\bfx}_2^2 \qquad \mbox{for every $\bfx\in\mathbb R^d$}
\]
with only $\tilde O(k/\eps^2)$ samples. This fact immediately follows from applying the more standard guarantee for \emph{leverage score sampling} on a concatenated matrix $[\bfA; \sqrt\lambda\bfI]\in\mathbb R^{(n+d)\times d}$, where $\bfI$ is the $d\times d$ identity.

For an $\ell_p$ version of this result, a recent result of \cite{WY2023c} shows an analysis of \emph{root leverage score sampling} for nearly optimal $\ell_p$ subspace embeddings for $p<2$, where the sampling probabilities are taken to be proportional to $\bftau_i(\bfA)^{p/2}$, i.e., the $(p/2)$-th roots of the leverage score $\bftau_i(\bfA)$. Although the row sample size is roughly $\tilde O(\eps^{-2}n^{1-p/2}d^{p/2})$, which depends on $n$, this result can be applied recursively for $O(\log\log n)$ applications to reduce the sample complexity down to $\tilde O(\eps^{-4/p}d)$, which is nearly linear in $d$ and thus nearly optimal. We show that a similar analysis in fact also works for $p>2$, where we can reduce the number of rows to $\tilde O(\eps^{-2}n^{1-2/p}d)$ in one round and $\tilde O(\eps^{-p}d^{p/2})$ rows by applying this result recursively.

With the $\ell_p$ subspace embedding theorem in hand, we can now apply a similar trick as \cite{CMM2017}. We set $\lambda = \norm*{\bfA-\bfA_k}_F^2/k$ and compute a subspace embedding for the matrix formed by concatenating $\bfA$ with $\sqrt\lambda\bfI$ via root leverage score sampling. This is equivalent to root ridge leverage score sampling for the matrix $\bfA$, whose scores sum to $k$. Thus, in each step of the recursive sampling, the number of rows sampled is $\tilde O(\eps^{-2} n^{1-p/2}k^{p/2})$ for $p<2$ and $\tilde O(\eps^{-2}n^{1-2/p}k)$ for $p>2$. After $O(\log \log n)$ steps of recursion, the final sample size is $\tilde O(\eps^{-4/p}k)$ rows for $p<2$ and $\tilde O(\eps^{-p} k^{p/2})$ rows for $p>2$. We note that coreset sizes we obtain in Theorems \ref{thm:strong-coreset-p<2} and \ref{thm:strong-coreset-p>2} are a direct result of this recursive sampling scheme. Furthermore, the resulting additive-multiplicative subspace embedding guarantee is
\[
    \norm*{\bfS\bfA\bfx}_p^p = (1\pm\eps)\norm*{\bfA\bfx}_p^p \pm \eps \lambda^{p/2}\norm*{\bfx}_p^p.
\]

Now, we can apply the above $\ell_p$ additive-multiplicative subspace embedding guarantees for the sampling matrix $\bfS$ on each column of $\norm*{\bfA\bfX\bfH}_{p,p}^p$ to obtain the approximation guarantee
\[
    \norm*{\bfS\bfA\bfX\bfH}_{p,p}^p = (1\pm\eps) \norm*{\bfA\bfX\bfH}_{p,p}^p \pm \eps \lambda^{p/2}\norm*{\bfX\bfH}_{p,p}^p.
\]
Now by applying Dvoretzky's theorem again to map the $(p,p)$-norm back to the $(p,2)$-norm, we obtain
\[
    \norm*{\bfS\bfA\bfX}_{p,2}^p = (1\pm\eps) \norm*{\bfA\bfX}_{p,2}^p \pm \eps \lambda^{p/2} \norm*{\bfX}_{p,2}^p.
\]
Finally, it remains to bound $\lambda^{p/2} \norm*{\bfX}_{p,2}^p$, but here we will encounter some problems.

\subsubsection{Problems when bounding the additive error}

To bound the additive error $\lambda^{p/2} \norm*{\bfX}_{p,2}^p$, we will case on $p<2$ and $p>2$. We may assume without loss of generality that $\bfX$ has at most $n$ rows, by restricting to the row span of $\bfA$. Then for $p<2$, $\lambda^{p/2}$ is at most
\begin{equation}\label{eq:lambda-bound-p<2}
    \lambda^{p/2} = \frac{\norm{\bfA-\bfA_k}_F^p}{k^{p/2}} \leq \frac{\norm{\bfA(\bfI - \bfP^*)}_F^{p}}{k^{p/2}} \leq \frac{\norm{\bfA(\bfI - \bfP^*)}_{p,2}^{p}}{k^{p/2}} = \frac{\OPT}{k^{p/2}}
\end{equation}
by the monotonicity of $\ell_p$ norms, while for $p>2$, $\lambda^{p/2}$ is at most
\begin{equation}\label{eq:lambda-bound-intro}
    \lambda^{p/2} = \frac{\norm*{\bfA-\bfA_k}_F^{p}}{k^{p/2}} \leq \frac{\norm*{\bfA(\bfI-\bfP^*)}_F^{p}}{k^{p/2}} \leq \frac{n^{p/2-1}\norm*{\bfA(\bfI-\bfP^*)}_{p,2}^{p}}{k^{p/2}} = \frac{n^{p/2-1}\OPT}{k^{p/2}}.
\end{equation}
Furthermore, 
\[
    \norm{\bfX}_{p,2}^p \leq \begin{cases}
        n^{1-p/2}\norm{\bfX}_{2,2}^p & \text{if $p<2$} \\
        \norm{\bfX}_{2,2}^p & \text{if $p>2$}
    \end{cases} \leq \begin{cases}
        n^{1-p/2}k^{p/2} & \text{if $p<2$} \\
        k^{p/2} & \text{if $p>2$}
    \end{cases}
\]
by using that $\rank(\bfX)\leq k$ and $\norm{\bfX}_2\leq 1$. Then overall, we obtain a bound of
\[
    \lambda^{p/2}\norm{\bfX}_{p,2}^p \leq \begin{dcases}
        n^{1-p/2} \OPT & \text{if $p<2$} \\
        n^{p/2-1} \OPT & \text{if $p>2$}
    \end{dcases}
\]
which are off by $\poly(n)$ factors from $\OPT$, which is the error that we can tolerate. In order to fix this problem and improve our analysis by $\poly(n)$ factors, we will use two different types of ``flattening'' tricks, one for $p<2$ and one for $p>2$, which we discuss in the next two sections.

\subsubsection{Idea 2: splitting rows for sharper additive error bounds for \texorpdfstring{$p<2$}{p<2}}

To overcome the previous issue for $p<2$, we will sharpen the bound of \eqref{eq:lambda-bound-p<2}. The loose bound that we will tighten is bounding the Frobenius norm loss $\norm{\bfA(\bfI-\bfP^*)}_F^p$ by the $(p,2)$-norm loss $\norm{\bfA(\bfI-\bfP^*)}_{p,2}^p$. For general matrices, this bound is indeed tight since the rows of $\bfA(\bfI-\bfP^*)$ could be imbalanced so that most of the mass is concentrated on a few rows. However, this bound is loose when the rows are flat, in which case there can be a $\poly(n)$ factor separation in the two quantities. We will show how to recover this separation.

A classic result of \cite{VX2012} shows that the sensitivity scores \eqref{eq:sensitivities} for $\ell_p$ subspace approximation sum to at most $O(k)$ for $p < 2$. Then, a standard flattening argument shows that by replacing rows $\bfa_i$ with large sensitivity with $l$ copies of the scaled row $\bfa_i/l^{1/p}$, we obtain a new matrix $\bfA'$ with $n' \leq 2n$ rows that are each just scaled copies of rows of $\bfA$, such that $\bfsigma_{i'}(\bfA') = O(k/n)$ for every row $i'\in[n']$ and $\norm{\bfA'(\bfI-\bfP_F)}_{p,2}^p = \norm{\bfA(\bfI-\bfP_F)}_{p,2}^p$ for every $F\in\mathcal F_k$. Because this matrix is now flat, it can be shown that
\[
    \norm{\bfA'(\bfI-\bfP^*)}_F^2 \lesssim \parens*{k/n}^{2/p-1}\OPT^{2/p}.
\]
Thus by replacing $\bfA$ with $\bfA'$, we obtain a matrix formed by the rows of $\bfA$ that gives the same objective function, yet has a much smaller additive error when bounding $\lambda$, giving
\[
    \lambda^{p/2} = \frac{\norm{\bfA'-\bfA_{k}'}_F^p}{k^{p/2}} \leq \frac{\norm{\bfA'(\bfI - \bfP^*)}_F^{p}}{k^{p/2}} \leq (k/n)^{1-p/2} \frac{\norm{\bfA'(\bfI - \bfP^*)}_{p,2}^{p}}{k^{p/2}} = (k/n)^{1-p/2}\frac{\OPT}{k^{p/2}}.
\]
rather than the original bound in \eqref{eq:lambda-bound-p<2}. We note, however, that this argument is still lossy, since the standard sensitivity-based flattening argument would flatten \emph{any} matrix, whereas we only need this result for a single constant factor approximate subspace $\tilde F$. Thus, we instead explicitly compute a constant factor bicriteria solution, which can be done very quickly \cite{DTV2011, FKW2021, WY2023b} (see Lemma \ref{lem:fast-sens}), and flatten this particular solution nearly optimally, so that we instead get the bound
\[
    \lambda^{p/2} = \frac{\norm{\bfA'-\bfA_{k}'}_F^p}{k^{p/2}} \leq \frac{\norm{\bfA'(\bfI - \bfP_{\tilde F})}_F^{p}}{k^{p/2}} \leq (1/n)^{1-p/2} \frac{\norm{\bfA'(\bfI - \bfP_{\tilde F})}_{p,2}^{p}}{k^{p/2}} = n^{p/2-1}\frac{O(\OPT)}{k^{p/2}}.
\]
Thus, we recover the extra factor of $n^{1-p/2}$ lost when converting from the $\ell_p$ norm to the $\ell_2$ norm. This completes our proof sketch for $p<2$.

\subsubsection{Idea 3: Dvoretzky's theorem for sharper additive error bounds for \texorpdfstring{$p>2$}{p>2}}

To improve our argument for $p>2$, we note that we have an additional degree of freedom when choosing to concatenate $\bfA$ with $\sqrt\lambda \bfI$ when analyzing the ridge leverage score sampling algorithm. Indeed, as long as we concatenate $\bfA$ with $\sqrt\lambda \bfU$ for any orthonormal matrix $\bfU$, then the leverage scores of $\bfA$ concatenated with $\sqrt\lambda\bfU$ will have leverage scores which coincide with the ridge leverage scores of $\bfA$, since
\[
    \bfa_i^\top(\bfA^\top\bfA + \lambda \bfU^\top\bfU)^{-1}\bfa_i = \bfa_i^\top(\bfA^\top\bfA + \lambda \bfI)^{-1}\bfa_i.
\]
The resulting guarantee is that
\begin{equation}\label{eq:add-mult-subspace-embedding}
    \norm*{\bfS\bfA\bfx}_p^p = (1\pm\eps)\norm*{\bfA\bfx}_p^p \pm \eps \lambda^{p/2}\norm*{\bfU\bfx}_p^p,
\end{equation}
so we may select $\bfU$ to be an orthonormal matrix which makes this additive error as small as possible. We will choose $\bfU$ to be a random $n\times d$ orthonormal matrix $\bfG$ scaled by $n^{-1/2}$, which has the advantage of flattening the mass of $\bfx$ and thus minimizing the $\ell_p$ norm.

By Dvoretzky's theorem \cite{Dvo1961, FLM1977, PVZ2017}, it follows that as long as $n$ is at least $\tilde O(k^{p/2})\poly(\eps^{-1}) = \tilde O(k^{p/2})\poly(\eps^{-1})$, then for any $\bfx$ in a fixed $k$-dimensional subspace, we will have that
\begin{equation}\label{eq:dvoretzky-reg}
    \frac1{n^{p/2}}\norm*{\bfG\bfx}_p^p = (1\pm\eps)n^{1-p/2}\norm*{\bfx}_2^p.
\end{equation}
This cancels out with the factor of $n^{p/2-1}$ that we lost in \eqref{eq:lambda-bound-intro}, giving us a sharp enough additive error. It may be tempting to reduce the additive error even further by choosing $\bfG$ to have $m\gg n$ rows rather than just $n$. However, this would affect the total number of rows sampled, since we would then need to oversample the root ridge leverage scores by a factor of $m^{p/2-1}$, which would increase the sample complexity. Note also that once the additive error is sufficiently small, \eqref{eq:add-mult-subspace-embedding} would give the purely multiplicative subspace embedding guarantee $\norm{\bfS\bfA\bfx}_p^p = (1\pm\eps)\norm{\bfA\bfx}_p^p$, for which there is a sample complexity lower bound of $\Omega(d^{p/2}/\eps)$ \cite{LWW2021}.

Although we have fixed the $n^{p/2-1}$ factor, we must now address a subtle issue. The above analysis works for a fixed rank $k$ subspace specified by the low rank matrix $\bfX$. However, if we want this guarantee for every rank $k$ matrix $\bfX$ with operator norm $1$ as we need, then we run into problems, since for any fixed embedding $\bfG$ of dimension only $\tilde O(k^{p/2})\poly(\eps^{-1})$, there exists a choice of $\bfX$ which causes \eqref{eq:dvoretzky-reg} to fail. To fix our final problem, we consider sampling $\poly(n)$ many random matrices $\bfU$ and use the fact that our sampling theorem will succeed for all of these $\bfU$ by a union bound, with only a $\poly\log n$ overhead in the sample complexity. Then for any fixed $\bfX$, the probability that the previous analysis fails for all of the copies of $\bfU$ is at most $\exp(-\poly(n))$, so this success probability is high enough to union bound over a discretization of all rank $k$ matrices $\bfX$. This completes our proof sketch for $p>2$.

\subsubsection{Technical summary}

Our analysis introduces major new ideas involving ridge leverage scores, giving the first understanding of their behavior for norms other than the very algebraic case of $p = 2$. We achieve low-error additive-multiplicative affine embeddings using ridge leverage scores for these norms via novel flattening techniques both for $p < 2$ and for $p > 2$. Our techniques significantly reduce the regularization term, which may be of independent interest given the wide applicability of ridge leverage scores. Our additive-multiplicative affine embeddings are then used to embed the $(p,2)$-norm of matrices of the form of a ``low rank'' component plus a ``small fixed'' component, which exactly captures the $\ell_p$ subspace approximation objective function. In particular, ridge leverage scores enable us to avoid resorting to representative subspaces which would yield suboptimal bounds.

\subsection{Corollaries}

\subsubsection{Streaming algorithms}

A simple corollary of our nearly optimal constructions for strong coresets is that we immediately obtain similar results in \emph{streaming} model of computation. In the streaming model, the rows $\bfa_i$ of the input matrix $\bfA$ arrive one at a time, and we wish to maintain a strong coreset for $\bfA$. In this setting, the classic \emph{merge-and-reduce} technique (see, e.g., \cite{BDMMUWZ2020} for a discussion) shows that a construction for a coreset of size $\tilde O(k^c)\poly(\eps^{-1})$ can be converted into a streaming implementation of size $\tilde O(k^c)\poly(\eps^{-1}\log n)$ by setting the accuracy parameter to $\eps' = \eps/\log n$ and composing the coreset construction in a binary tree fashion. Recent work of \cite{CWZ2023} shows that this argument can in fact be sharpened to a $\poly(\log\log n)$ factor overhead rather than $\poly(\log n)$, by first computing an online coreset (see Section \ref{sec:intro:online}), as we show in Section \ref{sec:streaming-coreset}. %

\subsubsection{Entrywise \texorpdfstring{$\ell_p$}{Lp} low rank approximation}

For $p<2$, our nearly optimal coresets for $\ell_p$ subspace approximation imply new algorithms for the related problem of \emph{entrywise $\ell_p$ low rank approximation}. 

\begin{Definition}
Let $\bfA\in\mathbb R^{n\times d}$ and let $k$ be a rank parameter. Let $1\leq p < \infty$. Then, the \emph{entrywise $\ell_p$ low rank approximation} problem is the problem of minimizing the objective function
\[
    \norm{\bfA-\bfX}_{p,p}^p = \sum_{i=1}^n\sum_{j=1}^d \abs*{(\bfA-\bfX)_{i,j}}^p
\]
among all rank $k$ matrices $\bfX\in\mathbb R^{n\times d}$.
\end{Definition}

This problem is another computationally difficult variant of the low rank approximation problem, and approximation algorithms and hardness have been studied in a long line of work \cite{SWZ2017, CGKLPW2017, DWZZR2019, MW2021, JLLMW2021, WY2023b}. The works of \cite{JLLMW2021, WY2023b} show that for $p<2$, if we multiply $\bfA$ on the right by a dense matrix $\bfG$ of $p$-stable random variables \cite{Nol2020} and then compute an $\ell_p$ subspace approximation coreset $\bfS$ of $\bfA\bfG$ of size $\tilde O(k)$, then there exists a rank $k$ matrix $\bfV$ such that
\[
    \norm{\bfA - \bfV\bfS\bfA}_{p,p}^p \leq \tilde O(k^{1/p-1/2})\min_{\rank(\bfX) \leq k}\norm{\bfA - \bfX}_{p,p}^p.
\]
Among subset selection algorithms, this approximation guarantee is nearly optimal \cite{MW2021}. Furthermore, because $\bfS$ is constructed based on sketching and coresets for $\ell_p$ subspace approximation, this algorithm can be implemented in streaming and distributed settings, and previously discussed. However, these prior results had drawbacks. The result of \cite{JLLMW2021} relied on the coreset construction of \cite{SW2018}, and thus does not give a true row subset, meaning that the subset selection lower bounds of \cite{MW2021} do not apply. In the work of \cite{WY2023b}, this idea was applied in the setting of online coresets, but their online coreset required a size of at least $k^4$, which resulted in a suboptimal approximation factor of at least $k^{4(1/p-1/2)}$. Our result fixes both of these problems, by giving a true row subset implementation of this result, as well as the first online coreset algorithm which selects $\tilde O(k)$ rows with a $\tilde O(k^{1/p-1/2})$ distortion. %

\subsection{Open directions}

The main natural direction left open by our work is to tighten the dependence on $\eps$ in the coreset size both in the upper bounds and lower bounds. Currently, the best known lower bound on the number of rows required is $\tilde\Omega(k/\eps^2)$ for $p<2$ and $\tilde\Omega(k/\eps^2 + k^{p/2}/\eps)$ for $p>2$ via a reduction to lower bounds for $\ell_p$ subspace embeddings \cite{LWW2021, WY2023b}, while we have a dependence of $\eps^{-\max\{4/p,p\}}$ in our upper bounds. 

\begin{Question}
How many rows are necessary and sufficient for strong coresets for $\ell_p$ subspace approximation as a function of both $k$ and $\eps$?
\end{Question}

Similar questions can also be asked for other guarantees for row subset selection for $\ell_p$ subspace approximation, all of which are well-understood for $p = 2$ but remain to be settled for $p\neq 2$.

\begin{Question}
\label{q:weak-coreset}
How many rows are necessary and sufficient for a weak coreset $\bfS$ such that $\tilde F \coloneqq \arg\min_{F\in\mathcal F_k}\norm{\bfS\bfA(\bfI-\bfP_F)}_p^p$ satisfies
\[
    \norm{\bfA(\bfI-\bfP_{\tilde F})}_p^p \leq (1+\eps)\min_{F\in\mathcal F_k}\norm{\bfA(\bfI-\bfP_F)}_p^p,
\]
as a function of both $k$ and $\eps$?
\end{Question}

\begin{Question}
\label{q:span-k-sol}
How many rows are necessary and sufficient for a subset $S\subseteq[n]$ such that the span of the rows in $S$ contains a $k$-dimensional subspace $\tilde F$ such that
\[
    \norm{\bfA(\bfI-\bfP_{\tilde F})}_p^p \leq (1+\eps)\min_{F\in\mathcal F_k}\norm{\bfA(\bfI-\bfP_F)}_p^p,
\]
as a function of both $k$ and $\eps$?
\end{Question}

\begin{Question}
\label{q:span-sol}
How many rows are necessary and sufficient for a subset $S\subseteq[n]$ such that the span $\tilde F$ of the rows in $S$ of dimension $\dim(S)$ satisfies
\[
    \norm{\bfA(\bfI-\bfP_{\tilde F})}_p^p \leq (1+\eps)\min_{F\in\mathcal F_k}\norm{\bfA(\bfI-\bfP_F)}_p^p,
\]
as a function of both $k$ and $\eps$?
\end{Question}

Our strong coreset immediately implies upper bounds to all three questions above, but it is possible to improve further in some of these cases. The guarantee in Question \ref{q:span-k-sol} was studied by \cite{DV2007, SV2012} for $p\neq 2$ with an efficient construction achieving an upper bound of $\tilde O(k^2\cdot (k/\eps)^{p+1})$ due to \cite{DV2007} and an inefficient construction achieving an upper bound of $\tilde O(k^2/\eps)$ due to \cite[Theorem 3.1]{SV2012} as well as $\tilde O(k/\eps^2)$ for $p = 1$ $\tilde O(k/\eps)$ for $p\in(1,2)$ due to \cite[Theorem 1.9]{WY2024}. The result \cite{DV2007} has a better dependence on $\eps$ than our strong coresets as well as those of \cite{HV2020, WY2023b}, and \cite{SV2012} has a better dependence on $\eps$ for all $p$ and a better dependence on $k$ for $p > 4$. In particular, it is an interesting question to achieve a nearly linear dependence in $k$ for all $p$, and to construct such a subset of rows in polynomial time.

\section{Preliminaries}

For a matrix $\bfA\in\mathbb R^{n\times d}$, we write $\bfA = \bfU\bfSigma\bfV^\top$ for the singular value decomposition (SVD) of $\bfA$. For a rank parameter $k$, we let $\bfSigma_k$ denote the matrix $\bfSigma$ with all but the top $k$ singular values zeroed out, $\bfSigma_{\setminus k}$ for the matrix $\bfSigma$ with all but the bottom $d-k$ singular values zeroed out, and $\bfA_k = \bfU\bfSigma_k\bfV^\top$ for the optimal rank $k$ approximation of $\bfA$ under the Frobenius norm. 

For a subspace $F$, we let $\bfV_F$ denote some orthonormal basis for the subspace.

\subsection{Facts about Gaussians}

\subsubsection{Dvoretzky's theorem}

A classic result of Dvoretzky and Milman \cite{Dvo1961, Mil1971} shows that a random subspace of a normed space is approximately Euclidean. We will need the following version of this result for $\ell_p$ norms:

\begin{Theorem}[Dvoretzky's theorem for $\ell_p$ norms \cite{FLM1977, PVZ2017}]
\label{thm:dvoretzky}
Let $1\leq p < \infty$ and $0 < \eps < 1/p$. Let $n\geq O(\max\{\eps^{-2}k, \eps^{-1}k^{p/2}\})$, and let $\bfG\in\mathbb R^{n\times k}$ be an i.i.d.\ random Gaussian matrix. Then,
\[
    \Pr\braces*{\mbox{for all $\bfx\in\mathbb R^k$, }\norm*{\bfG\bfx}_p^p = (1\pm\eps) n\norm*{\bfx}_2^p} \geq \frac23
\]
\end{Theorem}

\subsubsection{Random linear combinations}

The next lemma shows that the $(p,2)$-norm of a matrix $\bfX$ can be estimated by the $\ell_p$ norm of $\bfX\bfg$ for a Gaussian vector $\bfg$.

\begin{Lemma}
\label{lem:lp-lp2}
Let $\bfX\in\mathbb R^{n\times d}$. Fix a constant $p>0$. Then, for each $\delta\in(0,1)$,
\[
    \Pr_{\bfg\sim\mathcal N(0,\bfI_d)}\braces*{\delta\norm{\bfX\bfg}_p^p \lesssim \norm{\bfX}_{p,2}^p \lesssim \frac1{\delta^p} \norm{\bfX\bfg}_p^p} \geq 1-\delta.
\]
\end{Lemma}
\begin{proof}
Note that $\bfe_i^\top\bfX\bfg$ is distributed as a standard Gaussian with variance $\norm{\bfe_i^\top\bfX}_2^2$.

For the first bound, we have $\E\abs{\bfe_i^\top\bfX\bfg}^p \lesssim \norm{\bfe_i^\top\bfX}_2^p$, so by Markov's inequality, $\norm{\bfX\bfg}_p^p \lesssim (1/\delta)\norm{\bfX}_p^p$ with probability at least $1-\delta/2$.

For the second bound, note that $\abs{\bfe_i^\top\bfX\bfg} \gtrsim \delta\norm{\bfe_i^\top\bfX}_2$ with probability at least $1-\delta/4$. Then,
\[
    \E\bracks*{\sum_{i=1}^n \norm{\bfe_i^\top\bfX}_2^p \cdot \mathbbm{1}\braces*{\abs{\bfe_i^\top\bfX\bfg} \lesssim \delta\norm{\bfe_i^\top\bfX}_2}} \leq \frac{\delta}{4} \norm{\bfX}_{p,2}^p
\]
so by Markov's inequality, this is at most $\norm{\bfX}_{p,2}^p/2$ with probability at least $1-\delta/2$. Note also that
\begin{align*}
    \norm{\bfX\bfg}_p^p &\geq \sum_{i=1}^n \abs{\bfe_i^\top\bfX\bfg}^p \cdot \mathbbm{1}\braces*{\abs{\bfe_i^\top\bfX\bfg} \gtrsim \delta\norm{\bfe_i^\top\bfX}_2} \\
    &\gtrsim \sum_{i=1}^n \delta^p\norm{\bfe_i^\top\bfX}_2^p \cdot \mathbbm{1}\braces*{\abs{\bfe_i^\top\bfX\bfg} \gtrsim \delta\norm{\bfe_i^\top\bfX}_2} \\
    &= \sum_{i=1}^n \delta^p\norm{\bfe_i^\top\bfX}_2^p \cdot \parens*{1-\mathbbm{1}\braces*{\abs{\bfe_i^\top\bfX\bfg} \lesssim \delta\norm{\bfe_i^\top\bfX}_2}}
\end{align*}
Thus with probability at least $1-\delta/2$, we have that
\[
    \frac1{\delta^p} \norm{\bfX\bfg}_p^p \geq \norm{\bfX}_{p,2}^p - \sum_{i=1}^n \norm{\bfe_i^\top\bfX}_2^p \cdot \mathbbm{1}\braces*{\abs{\bfe_i^\top\bfX\bfg} \lesssim \delta\norm{\bfe_i^\top\bfX}_2} \geq \frac12 \norm{\bfX}_{p,2}^p.\qedhere
\]
\end{proof}

\subsection{Flattening}

It is known that constant factor bicriteria solutions for $\ell_p$ subspace approximation can be computed quickly via convex relaxations \cite{DTV2011} or by combining sketching techniques with $\ell_p$ Lewis weight sampling \cite{FKW2021, WY2023b}. The following lemma gives a version of \cite[Algorithm 3]{WY2023b} that is optimized for running time.

\begin{Lemma}[Fast constant factor approximation]
\label{lem:fast-sens}
Let $\bfA\in\mathbb R^{n\times d}$, $1\leq p \leq 2$, and $k\in\mathbb N$. Let $\bfG\in\mathbb R^{t\times d}$ be a sparse embedding matrix \cite{NN2013, Coh2016} with $t = O(k\log(n/\delta))$ and sparsity $s = O(\log(n/\delta))$. Let $\tilde F$ denote the span of $O(t\log(t/\delta))$ rows sampled according to the $\ell_p$ Lewis weights of $\bfA\bfG^\top$ \cite{CP2015}. Then, with probability at least $1-\delta$, the following hold:
\begin{itemize}
    \item $\norm{\bfA(\bfI-\bfP_{\tilde F})}_{p,2}^p \leq O(\OPT).$
    \item The subspace $\tilde F$ can be computed in $\tilde O(\nnz(\bfA) + t^\omega)$ time
\end{itemize}
\end{Lemma}
\begin{proof}
The correctness is shown in \cite{WY2023b}, so it remains to argue the running time. The sparse embedding matrix $\bfG$ only requires time $\tilde O(\nnz(\bfA)\log(1/\delta))$ to apply due to its sparsity. The $\ell_p$ Lewis weights of $\bfA\bfG^\top$ can then be computed in time $\tilde O(\nnz(\bfA\bfG^\top) + t^\omega) = \tilde O(\nnz(\bfA) + t^\omega)$ \cite{CP2015}.
\end{proof}

By using Lemma \ref{lem:fast-sens}, we will obtain a fast algorithm for quickly \emph{flattening} a matrix by splitting rows, which will be a crucial component of our sampling algorithm for $p<2$. Similar techniques have long been used in the literature of $\ell_p$ subspace embeddings \cite{BLM1989, CP2015, MMWY2022, WY2023a}.

\begin{Lemma}[Flattening]
\label{lem:flattening}
Let $\bfA\in\mathbb R^{n\times d}$, $1\leq p < \infty$, and $k\in\mathbb N$. Let $\tilde F\subseteq\mathbb R^d$ be a fixed subspace. Then, there is an $n'\times d$ matrix $\bfA'$ with $n \leq n' \leq (3/2) n$ such that $\norm{\bfA(\bfI-\bfP_F)}_{p,2}^p = \norm{\bfA'(\bfI-\bfP_F)}_{p,2}^p$ for every $F\in\mathcal F_k$ and 
\[
    \norm{\bfa_i'^\top(\bfI-\bfP_{\tilde F})}_2^p \leq \frac2n \norm{\bfA'(\bfI-\bfP_{\tilde F})}_{p,2}^p
\]
for every $i\in[n']$. Furthermore, the rows of $\bfA'$ are reweighted rows of $\bfA$.
\end{Lemma}
\begin{proof}
The proof follows, e.g., \cite[Lemma 2.10]{MMWY2022}. Note that if we replace a row $\bfa_i$ by $l$ copies of the scaled row $\bfa_i/l^{1/p}$, then $\norm{\bfA(\bfI-\bfP_F)}_{p,2}^p = \norm{\bfA'(\bfI-\bfP_F)}_{p,2}^p$ and for every row $i'$ in $\bfA'$ that is a copy of $\bfA$, $\norm{\bfa_{i'}'^\top(\bfI-\bfP_{\tilde F})}_2^p = \norm{\bfa_i^\top(\bfI-\bfP_{\tilde F})}_2^p / l$. Now for every row $i$ in $\bfA$ such that $\norm{\bfa_i^\top(\bfI-\bfP_{\tilde F})}_2^p \geq 2\norm{\bfA^\top(\bfI-\bfP_{\tilde F})}_{p,2}^p/n$, replace the row $\bfa_i$ with 
\[
    l_i \coloneqq \ceil*{\frac{\norm{\bfa_i^\top(\bfI-\bfP_{\tilde F})}_2^p/\norm{\bfA^\top(\bfI-\bfP_{\tilde F})}_{p,2}^p}{2/n}}
\]
copies of $\bfa_i / l_i^{1/p}$. Note then that the number of rows we add is at most
\[
    \sum_{i = 1}^n (l_i - 1) \leq \sum_{i = 1}^n \frac{\norm{\bfa_i^\top(\bfI-\bfP_{\tilde F})}_2^p/\norm{\bfA^\top(\bfI-\bfP_{\tilde F})}_{p,2}^p}{2/n} \leq \frac{n}{2}.
\]
Furthermore, by construction, every row in the new matrix $\bfA'$ satisfies
\[
    \norm{\bfa_i'^\top(\bfI-\bfP_{\tilde F})}_2^p \leq \frac2n \norm{\bfA'(\bfI-\bfP_{\tilde F})}_{p,2}^p.
\]
\end{proof}

The advantage of flattening is that for $p<2$, it makes the $\ell_2$ subspace approximation cost much smaller than the $\ell_p$ subspace approximation cost. We will exploit the following result later in our results for $p<2$.

\begin{Lemma}
\label{lem:flat-l2-lp}
Let $\bfA\in\mathbb R^{n\times d}$, $1\leq p \leq 2$, and $k\in\mathbb N$. Let $\tilde F\subseteq\mathbb R^d$ be a subspace. Suppose that 
\[
    \norm{\bfa^\top_i(\bfI-\bfP_{\tilde F})}_2^p \leq \frac{C}{n}\norm{\bfA(\bfI-\bfP_{\tilde F})}_{p,2}^p
\]
for every $i\in[n]$. Then, we have
\[
    \norm{\bfA(\bfI-\bfP_{\tilde F})}_F \leq (C/n)^{1/p-1/2} \norm{\bfA(\bfI-\bfP_{\tilde F})}_{p,2}.
\]
\end{Lemma}
\begin{proof}
We have
\begin{align*}
    \norm{\bfA(\bfI-\bfP_{\tilde F})}_F^2 &= \sum_{i=1}^{n} \norm{\bfa^\top_i(\bfI-\bfP_{\tilde F})}_2^2 = \sum_{i=1}^{n} \norm{\bfa^\top_i(\bfI-\bfP_{\tilde F})}_2^p (\norm{\bfa^\top_i(\bfI-\bfP_{\tilde F})}_2^p)^{2/p-1} \\
    &\leq \sum_{i=1}^{n} \norm{\bfa^\top_i(\bfI-\bfP_{\tilde F})}_2^p \parens*{\frac{C}{n}\norm{\bfA(\bfI-\bfP_{\tilde F})}_{p,2}^p}^{2/p-1} \\
    &= \parens*{C/n}^{2/p-1}\norm{\bfA(\bfI-\bfP_{\tilde F})}_{p,2}^{2}.\qedhere
\end{align*}
\end{proof}

\subsection{Properties of ridge leverage scores}

It is known that for $\lambda = \norm{\bfA-\bfA_k}_F^2/k$, the ridge leverage scores have a small sum.

\begin{Lemma}[Sum of ridge leverage scores \cite{CMM2017}]
\label{lem:sum-of-rls}
Let $\lambda = \norm*{\bfA-\bfA_k}_F^2 / k$. Then,
\[
    \sum_{i=1}^n \bftau_i^\lambda(\bfA) \leq 2k
\]
\end{Lemma}

Next, we show that ridge leverage scores upper bound the $\ell_2$ subspace approximation sensitivities \eqref{eq:sensitivities}. 

\begin{Lemma}[Ridge leverage scores bound sensitivities]
\label{lem:ridge-leverage-score-sensitivity-bound}
Let $\lambda = \norm*{\bfA-\bfA_k}_F^2 / k$. Then for every $i\in[n]$,
\[
    \bftau_i^\lambda(\bfA) \geq \frac1{48}\sup_{F\in\mathcal F_k}\frac{\norm*{\bfa_i^\top(\bfI-\bfP_F)}_2^2}{\norm*{\bfA(\bfI-\bfP_F)}_F^2}
\]
\end{Lemma}
\begin{proof}
Note that
\[
    \norm*{\bfA-\bfA_{2k}}_2^2 = \bfsigma_{k+1}^2(\bfA-\bfA_k) \leq \frac1k \sum_{j=1}^k \bfsigma_{j}^2(\bfA-\bfA_k) \leq \frac{\norm{\bfA-\bfA_k}_F^2}{k} = \lambda
\]
so
\begin{align*}
    \bftau_i^\lambda(\bfA) &= \sup_{\bfx\in\mathbb R^d} \frac{[\bfA\bfx](i)^2}{\norm{\bfA\bfx}_2^2 + \lambda\norm{\bfx}_2^2} \\
    &= \sup_{\bfx\in\mathbb R^d} \frac{[\bfA\bfx](i)^2}{\norm{\bfA_{2k}\bfx}_2^2 + \norm{(\bfA-\bfA_{2k})\bfx}_2^2 + \lambda\norm{\bfx}_2^2} \\
    &\geq \sup_{\bfx\in\mathbb R^d} \frac{[\bfA\bfx](i)^2}{\norm{\bfA_{2k}\bfx}_2^2 + 2\lambda\norm{\bfx}_2^2}.
\end{align*}
Now let $F\in\mathcal F_k$ be any rank $k$ subspace. Let $G$ denote the span of the rows of $\bfA_{2k}$, $F$, and $\bfa_i$, which is a subspace of dimension at most $3k+1$. We then set $\bfx = \bfP_G(\bfI-\bfP_F)\bfg$ for a standard normal Gaussian vector $\bfg$. Note then that
\[
    [\bfA\bfx](i) = \bfa_i^\top\bfP_G(\bfI-\bfP_F)\bfg = \bfa_i^\top(\bfI-\bfP_F)\bfg
\]
is distributed as a Gaussian with variance $\norm{\bfa_i^\top(\bfI-\bfP_F)}_2^2$, so
\[
    \Pr\braces*{[\bfA\bfx](i)^2 \geq \norm{\bfa_i^\top(\bfI-\bfP_F)}_2^2/3} > \frac12.
\]
Note also that
\[
    \E\bracks*{\norm{\bfA_{2k}\bfx}_2^2} = \E\bracks*{\norm{\bfA_{2k}\bfP_G(\bfI-\bfP_F)\bfg}_2^2} = \E\bracks*{\norm{\bfA_{2k}(\bfI-\bfP_F)\bfg}_2^2} \leq \norm{\bfA(\bfI-\bfP_F)}_F^2
\]
and
\[
    \E[\lambda\norm{\bfx}_2^2] = \E[\lambda\norm{\bfP_G(\bfI-\bfP_F)\bfg}_2^2] \leq \lambda (3k+1) \leq 4\norm{\bfA-\bfA_k}_F^2.
\]
Then by Markov's inequality, we have
\[
    \Pr\braces*{\norm{\bfA_{2k}\bfx}_2^2 + 2\lambda\norm{\bfx}_2^2 \leq 16\norm{\bfA(\bfI-\bfP_F)}_F^2} \geq \frac12.
\]
Thus with positive probability, there exists a vector $\bfx$ such that
\[
    \bftau_i^\lambda(\bfA) \geq \frac{[\bfA\bfx](i)^2}{\norm{\bfA_{2k}\bfx}_2^2 + 2\lambda\norm{\bfx}_2^2} \geq \frac1{48}\frac{\norm{\bfa_i^\top(\bfI-\bfP_F)}_2^2}{\norm{\bfA(\bfI-\bfP_F)}_F^2}.
\]
Since $F$ was arbitrary, we conclude as desired.
\end{proof}

\section{Main sampling theorems}

\subsection{Affine ridge embedding for the \texorpdfstring{$(p,2)$}{(p,2)}-norm}

We first show a simple lemma which states that $\bftau_i^\lambda(\bfA)$ upper bounds the first $n$ leverage scores of $[\bfA:\sqrt\lambda\bfU]$ for any approximately orthogonal $\bfU$.

\begin{Lemma}
Let $\bfA\in\mathbb R^{n\times d}$ and $\lambda>0$. Let $\bfU\in\mathbb R^{n'\times d}$ satisfy $\bfU^\top\bfU \succeq \bfI/C$ and let $\bfA' = [\bfA; \sqrt{C\lambda}\bfU]$. Then,
\[
    \bftau_i^\lambda(\bfA) \geq \bftau_i(\bfA').
\]
\end{Lemma}
\begin{proof}
We have
\begin{align*}
\bftau_i^\lambda(\bfA) &= \sup_{\bfx\in\mathbb R^d} \frac{[\bfA\bfx](i)^2}{\norm{\bfA\bfx}_2^2 + \lambda \norm{\bfx}_2^2} \geq \sup_{\bfx\in\mathbb R^d} \frac{[\bfA\bfx](i)^2}{\norm{\bfA\bfx}_2^2 + C\lambda \norm{\bfU\bfx}_2^2} = \bftau_i(\bfA').
\end{align*}
\end{proof}

We now prove the following affine root ridge leverage score sampling lemma for the $(p,2)$-norm, which generalizes a root leverage score sampling theorem (Theorem \ref{thm:root-leverage-score-sampling-affine}) for vectors to matrices and incorporates ridge regularization. The main workhorse behind this lemma is Theorem \ref{thm:root-leverage-score-sampling-affine}, which establishes a relative error $\ell_p$ affine embedding theorem for root ridge leverage score sampling, and generalizes recent work of \cite{WY2023c} by handling the case of $p>2$ as well as allowing for an affine translation rather than just subspaces. 

\begin{Lemma}
\label{lem:root-ridge-affine}
Let $1\leq p <\infty$. Let $\alpha \leq O(\eps^2) / ((\log n)^3 + \log(1/\delta))$. Let $\bfS$ be the $\ell_p$ sampling matrix with probabilities $\{q_i\}_{i=1}^n$ for
\[
    q_i \geq \begin{cases}
    \min\braces*{1,n^{p/2-1}\bftau_i^\lambda(\bfA)^{p/2}/\alpha} & \text{if $p>2$} \\
    \min\braces*{1,\bftau_i^\lambda(\bfA)^{p/2}/\alpha} & \text{if $p<2$} \\
    \end{cases}
\]
Let $\bfB\in\mathbb R^{n\times d}$ be such that
\[
    \frac{\norm{\bfe_i^\top\bfB}_2^p}{\norm{\bfB}_{p,2}^p} \leq \begin{cases}
    \min\braces*{1,n^{p/2-1}\bftau_i^\lambda(\bfA)^{p/2}} & \text{if $p>2$} \\
    \min\braces*{1,\bftau_i^\lambda(\bfA)^{p/2}} & \text{if $p<2$} \\
    \end{cases}
\]
for each $i\in[n]$. Fix $\bfU\in\mathbb R^{n'\times d}$ satisfy $\bfU^\top\bfU \succeq \bfI/C$. Then,
\[
    \norm{\bfS(\bfA\bfX + \bfB)}_{p,2}^p = (1\pm\eps)\norm{\bfA\bfX + \bfB}_{p,2}^p \pm \eps\parens*{\norm{C\sqrt\lambda \bfU\bfX}_{p,2}^p + O(\log(nd/\eps\delta))^{p/2}\norm{\bfB}_{p,2}^p}.
\]
\end{Lemma}
\begin{proof}
Let $m \geq O(\max\{\eps^{-2} d, \eps^{-1}d^{p/2}\})$ and let $\bfG\in\mathbb R^{d\times m}$ be drawn with standard Gaussian entries. For each $i\in[n]$ and $j\in[m]$, $\bfe_i^\top\bfB\bfG\bfe_j$ is distributed as a Gaussian with variance $\norm{\bfe_i^\top\bfB}_2^2$. Thus, for each $i\in[n]$ and $j\in[m]$,
\[
    \Pr_\bfG\braces*{\abs{\bfe_i^\top\bfB\bfG\bfe_j} \leq O(\sqrt{\log(nm/\delta)})\norm{\bfe_i^\top\bfB}_2} \geq 1 - \frac{\delta}{2nm}.
\]
By a union bound, this bound is simultaneously true for all $i\in[n]$ and $j\in[m]$ with probability at least $1-\delta/2$. Conditioned on this event, we have $\norm{\bfB\bfG\bfe_j}_p \leq R$ for every $j\in[m]$ for $R = O(\sqrt{\log(nm/\delta)})\norm{\bfB}_{p,2}$. 

We will now apply root leverage score sampling (Theorem \ref{thm:root-leverage-score-sampling-affine}) $m$ times, once for each $j\in[m]$, with the following parameters:
\begin{itemize}
    \item set matrix $\bfA$ to the $(n+n')\times d$ matrix $[\bfA; \sqrt{C\lambda} \bfU]$
    \item set affine vector $\bfb$ to the $(n+n')$-dimensional vector $[\bfB\bfG\bfe_j; 0]$ with $n'$ trailing zeros
    \item $R = O(\sqrt{\log(nm/\delta)})\norm{\bfB}_{p,2}$
    \item failure probability $\delta/2m$
    \item sampling matrix which samples the first $n$ rows with $\bfS$ and the last $n'$ rows with probability $1$
\end{itemize}
By a union bound, the sampling guarantee succeeds for all $j\in[m]$ with probability at least $1-\delta/2$. This gives the following guarantee for every $j\in[m]$ and every $\bfX\in\mathbb R^{d\times d}$
\[
    \norm{\bfS(\bfA\bfX\bfG\bfe_j + \bfB\bfG\bfe_j)}_p^p + \norm{\sqrt{C\lambda} \bfU\bfX\bfG\bfe_j}_p^p = (1\pm\eps)\norm{\bfA\bfX\bfG\bfe_j + \bfB\bfG\bfe_j}_p^p + (1\pm\eps)\norm{\sqrt{C\lambda} \bfU\bfX\bfG\bfe_j}_p^p \pm \eps R^p
\]
By subtracting $\norm{\sqrt{C\lambda} \bfU\bfX\bfG\bfe_j}_p^p$ from both sides, we obtain
\[
    \norm{\bfS(\bfA\bfX\bfG\bfe_j + \bfB\bfG\bfe_j)}_p^p = (1\pm\eps)\norm{\bfA\bfX\bfG\bfe_j + \bfB\bfG\bfe_j}_p^p \pm \eps\norm{\sqrt{C\lambda} \bfU\bfX\bfG\bfe_j}_p^p \pm \eps R^p
\]
Finally, summing this guarantee over $j\in[m]$ and scaling by $1/m$ gives
\[
    \frac1m\norm{\bfS(\bfA\bfX + \bfB)\bfG}_{p,p}^p = (1\pm\eps)\frac1m\norm{(\bfA\bfX + \bfB)\bfG}_{p,p}^p \pm \eps\frac1m\norm{\sqrt{C\lambda} \bfU\bfX\bfG}_{p,p}^p \pm \eps R^p
\]
By Dvoretzky's theorem (Theorem \ref{thm:dvoretzky}) and rescaling $\eps$ by constant factors, we thus obtain
\[
    \norm{\bfS(\bfA\bfX + \bfB)}_{p,2}^p = (1\pm\eps)\norm{\bfA\bfX + \bfB}_{p,2}^p \pm \eps\norm{\sqrt{C\lambda} \bfU\bfX}_{p,2}^p \pm \eps R^p.\qedhere
\]
\end{proof}

\subsection{Results for \texorpdfstring{$p>2$}{p>2}}

We first give an error bound after one round of root ridge leverage score sampling.
\begin{Theorem}
\label{thm:main-rls-sampling-p>2-1-round}
Let $p>2$. Let $\bfA\in\mathbb R^{n\times d}$ with $n \geq n'$ for some $n' = O(k^{p/2}/\eps)$. Let $\alpha = \Theta(\eps^{2}) / ((\log n)^3 + \log(1/\delta))$.  Let $\bfS$ be the $\ell_p$ sampling matrix with probabilities $\{q_i\}_{i=1}^n$ for
\[
    q_i \geq \min\braces*{1,n^{p/2-1}\bftau_i^\lambda(\bfA)^{p/2}/\alpha}
\]
with $\lambda = \norm*{\bfA-\bfA_k}_F^2 / k$. Then, with probability at least $1-\delta$, for every $F\in\mathcal F_k$,
\[
    \norm*{\bfS\bfA(\bfI-\bfP_F)}_{p,2}^p = (1\pm\eps)\norm*{\bfA(\bfI-\bfP_F)}_{p,2}^p \pm O(\eps) O(\log(n/\eps\delta))^{p/2} \OPT.
\]
\end{Theorem}
\begin{proof}
Note that we can write
\[
    \bfA(\bfI-\bfP_F) = \bfA(\bfP^*-\bfP_F) + \bfA(\bfI-\bfP^*)
\]
Let $\bfX_F = \bfP^*-\bfP_F$ and let $\bfB = \bfA(\bfI-\bfP^*)$, so that $\bfA(\bfI-\bfP_F) = \bfA\bfX_F + \bfB$. Note that $\bfX_F$ is a matrix of rank at most $2k$ and operator norm at most $2$. Furthermore the sampling probabilities bound $\bfB$ as required by Lemma \ref{lem:root-ridge-affine} by Lemma \ref{lem:ridge-leverage-score-sensitivity-bound}.

Consider a net $N$ of size $n^{O(dk)}$ over the space of $d\times d$ matrices $\bfX$ with rank at most $2k$ and operator norm at most $2$, such that for every rank $2k$ $\bfX$ with operator norm at most $2$, there is $\bfX'\in N$ such that $\norm{\bfX-\bfX'}_2 \leq 1/n^{O(1)}$.

Let $\bfU\in\mathbb R^{2n\times d}$ be a random Gaussian matrix scaled by $1/\sqrt{2n}$. It is known that $\bfI/2 \preceq \bfU^\top\bfU \preceq 2\bfI$ with probability at least $1 - \exp(-\Omega(n))$ \cite[Theorem 1.1, Proposition 2.3]{RV2009}. Now consider drawing $U = O(\log(\abs{N}/\delta)) = O(dk)\log n + O(\log(1/\delta))$ such random matrices. Since $n \geq O(\log(n/\delta))$, we have by a union bound that all of the draws $\bfU$ satisfy the spectral bound with probability at least $1-\delta$. Similarly, the affine ridge sampling lemma (Lemma \ref{lem:root-ridge-affine}) succeeds for all draws of $\bfU$ with probability at least $1-\delta$.

Now for each fixed $\bfX'\in N$, let $\bfX' = \bfR\bfQ$ where $\bfQ$ has orthonormal rows and $\bfR$ is $n\times 2k$ with operator norm at most $2$. Then for each draw of $\bfU$, with probability at least $1/10$, we have that
\begin{align*}
    \norm{\bfU\bfX'}_{p,2}^p &= \norm{\bfU\bfR}_{p,2}^p \\
    &\lesssim \norm{\bfU\bfR\bfg}_{p}^p && \text{Lemma \ref{lem:lp-lp2}} \\
    &\lesssim n^{1-p/2}\norm{\bfR\bfg}_{2}^p && \text{Theorem \ref{thm:dvoretzky}} \\
    &\lesssim n^{1-p/2}\norm{\bfg}_{2}^p \lesssim n^{1-p/2} k^{p/2}
\end{align*}
where each inequality occurs with large enough constant probability. Thus, one of the $U$ draws of $\bfU$ satisfies this bound with probability at least $\delta/\abs{N}$. By a union bound, this is true simultaneously for ever $\bfX'\in N$. Then, for any $\bfX_F$, if $\bfX'$ is the closest matrix to $\bfX_F$ in $N$, then there exists a draw $\bfU$ such that
\[
    \norm{\bfU\bfX_F}_{p,2}^p \leq (\norm{\bfU\bfX'}_{p,2} + \norm{\bfU(\bfX_F-\bfX')}_{p,2})^p \lesssim n^{1-p/2} k^{p/2}.
\]
For this $\bfU$, we have
\begin{align*}
    \norm{\bfS\bfA(\bfI-\bfP_F)}_{p,2}^p &= \norm{\bfS(\bfA\bfX_F + \bfB)}_{p,2}^p \\
    &= (1\pm\eps)\norm{\bfA\bfX_F+\bfB}_{p,2}^p \pm O(\eps)\parens*{\norm{\sqrt\lambda \bfU\bfX_F}_{p,2}^p + O(\log(n/\eps\delta))^p \norm{\bfB}_{p,2}^p} && \text{Lemma \ref{lem:root-ridge-affine}} \\
    &= (1\pm\eps)\norm{\bfA(\bfI-\bfP_F)}_{p,2}^p \pm O(\eps)\parens*{\lambda^{p/2}n^{1-p/2}k^{p/2} + O(\log(n/\eps\delta))^p \OPT}
\end{align*}
Note that
\begin{equation}\label{eq:lambda-bound}
    \lambda^{p/2} = \frac{\norm*{\bfA-\bfA_k}_F^p}{k^{p/2}} \leq \frac{\norm*{\bfA(\bfI-\bfP^*)}_F^p}{k^{p/2}} \leq n^{p/2-1}\frac{\norm*{\bfA(\bfI-\bfP^*)}_{p,2}^p}{k^{p/2}} = \frac{n^{p/2-1}}{k^{p/2}}\OPT
\end{equation}
so we have the desired guarantee.
\end{proof}

Finally, we show that by applying Theorem \ref{thm:main-rls-sampling-p>2-1-round} recursively for $O(\log\log n)$ rounds, we obtain our desired sampling theorem. We will use the following recurrence:

\begin{Lemma}[Lemma 6.12, \cite{MMWY2022}]
\label{lem:recurrence}
Suppose $(a_i)_{i=0}^\infty$ satisfies the recurrence $a_{i+1} = \lambda a_i + b$ for some $b>0$ and $\lambda\in(0,1)$. Then,
\[
    a_i = \frac1{1-\lambda}\parens*{b - \lambda^i(b - (1-\lambda)a_0)}.
\]
\end{Lemma}

We then arrive at our main theorem for $p>2$. Note that Theorem \ref{thm:strong-coreset-p>2} follows by applying this result for $O(\log^* n)$ iterations, which reduces $n$ to $\poly(k/\eps)$.

\begin{Theorem}
\label{thm:main-rls-sampling-p>2}
Let $p>2$. Let $\bfA\in\mathbb R^{n\times d}$. There is an algorithm that runs in time $\tilde O(\nnz(\bfA) + d^\omega)$ time to construct a diagonal matrix $\bfS$ with
\[
    \nnz(\bfS) = \frac{O(k^{p/2})}{\eps^{p}}\bracks*{(\log n)^{3p/2} + (\log(1/\delta))^{p/2}}(\log(n/\delta))^{p^2/2}(\log\log n)^{p} = \frac{k^{p/2}}{\eps^{p}}(\log(n/\delta))^{O(p^2)}
\]
that satisfies Definition \ref{def:strong-coreset} with probability at least $1-\delta$.
\end{Theorem}
\begin{proof}
We will apply Theorem \ref{thm:main-rls-sampling-p>2-1-round} for $r = O(\log\log n)$ rounds, with $\eps$ set to $\eps/O(r\log(n/\eps\delta))^p$ and $\delta$ set to $\delta/r$. Let $\alpha$ and $s$ be the values given by Theorem \ref{thm:main-rls-sampling-p>2-1-round} with this setting of parameters. We first analyze the number of rows sampled at each round in expectation. Note first that if $n^{p/2-1}\bftau_i^\lambda(\bfA) \geq 1$, then $\bftau_i^\lambda(\bfA) \geq n^{2/p-1}$ so there are at most $O(k)n^{1-2/p}$ such rows $i\in[n]$, since $\bftau_i^\lambda(\bfA)$ sum to at most $2k$ by Lemma \ref{lem:sum-of-rls}. On the other hand, if $n^{p/2-1}\bftau_i^\lambda(\bfA)^{p/2}$, then $\bftau_i^\lambda(\bfA) \leq n^{2/p-1}$ so we have that
\[
    \sum_{i : n^{p/2-1}\bftau_i^\lambda(\bfA)^{p/2} \leq 1}n^{p/2-1}\bftau_i^\lambda(\bfA)^{p/2} \leq n^{p/2-1} (n^{2/p-1})^{p/2-1}\sum_{i : n^{p/2-1}\bftau_i^\lambda(\bfA)^{p/2} \leq 1}\bftau_i^\lambda(\bfA) \leq n^{1-2/p}\cdot 2k.
\]
Thus, in either case, we have
\[
    \sum_{i=1}^n \min\{1, n^{p/2-1}\bftau_i^\lambda(\bfA)^{p/2}\} \leq O(k)n^{1-2/p}.
\]
Thus, the expected number of sampled rows is at most $O(k)n^{1-2/p}/\alpha$. By Chernoff bounds, if the expected number of sampled rows is at least $O(\log(r/\delta))$, then with probability at least $1-\delta/r$, the number of sampled rows is within a constant factor of the expectation. Then by a union bound, for the first $r$ rounds of the recursive calls, we succeed in obtaining a $(1\pm\eps/r)$ approximation and reduce the number of rows from $m$ to $O(k)m^{1-2/p}/\alpha$. We now define $a_i$ to be the logarithm of the number of rows after the $i$th recursive call. Then,
\[
    a_{i+1} = (1-2/p) a_i + \log(O(s)/\alpha)
\]
so by Lemma \ref{lem:recurrence}, we have that
\[
    a_r = \frac{p}2\parens*{\log(O(k)/\alpha) - (1-2/p)^i (\log(O(k)/\alpha) - (2/p))(\log n)}
\]
so the number of rows is at most
\[
    \exp(a_r) = \parens*{\frac{O(k)}{\alpha}}^{p/2} = \frac{O(k^{p/2})}{\alpha^{p/2}} = \frac{O(k^{p/2})}{\eps^{p}}\bracks*{(\log n)^{3p/2} + (\log(1/\delta))^{p/2}}(\log(n/\delta))^{p^2/2}(\log\log n)^{p} .\qedhere
\]
\end{proof}

\subsection{Results for \texorpdfstring{$p<2$}{p < 2}}

The next theorem gives an error bound after one round of root ridge leverage score sampling. For $p<2$, we rely on a constant factor approximate solution $\tilde F$, which we will need to compute explicitly. Thus, we allow for the use of a slightly larger rank $k'\geq k$ to facilitate efficient computations of $\tilde F$.

\begin{Theorem}
\label{thm:main-rls-sampling-p<2-1-round}
Let $p<2$. Let $\alpha = \Theta(\eps^{2}) / ((\log n)^3 + \log(1/\delta))$.  Let $\bfS$ be the $\ell_p$ sampling matrix with probabilities $\{q_i\}_{i=1}^n$ for
\[
    q_i \geq \min\braces*{1,\bftau_i^\lambda(\bfA)^{p/2}/\alpha}
\]
with $\lambda = \norm*{\bfA-\bfA_{k'}}_F^2 / k'$ for $k'\geq k$. Furthermore, suppose that there is a rank $k'$ subspace $\tilde F$ such that
\[
    \norm{\bfa_i^\top(\bfI-\bfP_{\tilde F})}_2^p \leq O(1/n) \norm{\bfA(\bfI-\bfP_{\tilde F})}_{p,2}^p
\]
and $\norm{\bfA(\bfI-\bfP_{\tilde F})}_{p,2}^p \leq O(\OPT)$. Then, with probability at least $1-\delta$, for every $F\in\mathcal F_k$,
\[
    \norm*{\bfS\bfA(\bfI-\bfP_F)}_{p,2}^p = (1\pm\eps)\norm*{\bfA(\bfI-\bfP_F)}_{p,2}^{p/2}\pm O(\eps) O(\log(n/\eps\delta))^p \OPT.
\]
\end{Theorem}
\begin{proof}
Note that we can write
\[
    \bfA(\bfI-\bfP_F) = \bfA(\bfP^*-\bfP_F) + \bfA(\bfI-\bfP^*)
\]
Let $\bfX_F = \bfP^*-\bfP_F$ and let $\bfB = \bfA(\bfI-\bfP^*)$, so that $\bfA(\bfI-\bfP_F) = \bfA\bfX_F + \bfB$. Note that $\bfX_F$ is a matrix of rank at most $2k$ and operator norm at most $2$. Furthermore the sampling probabilities bound $\bfB$ as required by Lemma \ref{lem:root-ridge-affine} by Lemma \ref{lem:ridge-leverage-score-sensitivity-bound}.

We then apply Lemma \ref{lem:root-ridge-affine} with $\bfU = \bfI$, so that
\begin{align*}
    \norm{\bfS\bfA(\bfI-\bfP_F)}_{p,2}^p &= \norm{\bfS(\bfA\bfX_F + \bfB)}_{p,2}^p \\
    &= (1\pm\eps)\norm{\bfA\bfX_F+\bfB}_{p,2}^p \pm O(\eps)\parens*{\norm{\sqrt\lambda \bfX_F}_{p,2}^p + O(\log(n/\eps\delta))^p \norm{\bfB}_{p,2}^p} && \text{Lemma \ref{lem:root-ridge-affine}} \\
    &= (1\pm\eps)\norm{\bfA(\bfI-\bfP_F)}_{p,2}^p \pm O(\eps)\parens*{\lambda^{p/2}n^{1-p/2}k^{p/2} + O(\log(n/\eps\delta))^p \OPT}
\end{align*}
Note that
\begin{equation}\label{eq:lambda-bound-p<2-actual}
    \lambda^{p/2} \leq \frac{\norm*{\bfA-\bfA_{k'}}_F^p}{k'^{p/2}} \leq \frac{\norm*{\bfA(\bfI-\bfP_{\tilde F})}_F^p}{k'^{p/2}} \leq O(1/n)^{1-p/2}\frac{\norm*{\bfA(\bfI-\bfP_{\tilde F})}_{p,2}^p}{k'^{p/2}} = \frac{O(\OPT)}{k'^{p/2}n^{1-p/2}}
\end{equation}
so we have the desired guarantee.
\end{proof}

Finally, we show that by applying Theorem \ref{thm:main-rls-sampling-p<2-1-round} recursively for $O(\log\log n)$ rounds, we arrive at our main theorem for $p<2$. Note that Theorem \ref{thm:strong-coreset-p<2} follows by applying this result for $O(\log^* n)$ iterations, which reduces $n$ to $\poly(k/\eps)$.

\begin{Theorem}
\label{thm:main-rls-sampling-p<2}
Let $p<2$. Let $\bfA\in\mathbb R^{n\times d}$. There is an algorithm that runs in time $\tilde O(\nnz(\bfA) + d^\omega)$ time to construct a diagonal matrix $\bfS$ with
\[
    \nnz(\bfS) = \frac{O(k)}{\eps^{4/p}}\bracks*{(\log n)^{6/p+1} + (\log(1/\delta))^{2/p+1}}(\log(n/\delta))^{2}(\log\log n)^{4/p} = \frac{k}{\eps^{4/p}}(\log(n/\delta))^{O(1)}
\]
that satisfies Definition \ref{def:strong-coreset} with probability at least $1-\delta$.
\end{Theorem}
\begin{proof}
We will apply Theorem \ref{thm:main-rls-sampling-p<2-1-round} for $r = O(\log\log n)$ rounds, with $\eps$ set to $\eps/O(r\log(n/\eps\delta))^p$ and $\delta$ set to $\delta/r$. In order to satisfy the precondition of the existence of a ``flat'' solution, we quickly compute a constant factor solution via Lemma \ref{lem:fast-sens} and flatten via Lemma \ref{lem:flattening}, so that our condition is satisfied as long as we take $k' = O(k\log(n/\delta))$.

Let $\alpha$ be the value given by Theorem \ref{thm:main-rls-sampling-p<2-1-round} with this setting of parameters. We first analyze the number of rows sampled at each round in expectation. Since the ridge leverage scores sum to at most $2k'$ by Lemma \ref{lem:sum-of-rls}, we have that
\[
    \sum_{i=1}^n \bftau_i^\lambda(\bfA)^{p/2} \leq n^{1-p/2}\parens*{\sum_{i=1}^n \bftau_i^\lambda(\bfA)}^{p/2} = O(k'^{p/2}n^{1-p/2})
\]
by H\"older's inequality. Thus, the expected number of sampled rows is at most $O(k'^{p/2}n^{1-p/2})/\alpha$. By Chernoff bounds, if the expected number of sampled rows is at least $O(\log(r/\delta))$, then with probability at least $1-\delta/r$, the number of sampled rows is within a constant factor of the expectation. Then by a union bound, for the first $r$ rounds of the recursive calls, we succeed in obtaining a $(1\pm\eps/r)$ approximation and reduce the number of rows from $m$ to $O(k'^{p/2}m^{1-p/2})/\alpha$. We now define $a_i$ to be the logarithm of the number of rows after the $i$th recursive call. Then,
\[
    a_{i+1} = (1-p/2) a_i + \log(O(k'^{p/2})/\alpha)
\]
so by Lemma \ref{lem:recurrence}, we have that
\[
    a_r = \frac2p\parens*{\log(O(k'^{p/2})/\alpha) - (1-p/2)^i (\log(O(k'^{p/2})/\alpha) - (p/2))(\log n)}
\]
so the number of rows is at most
\[
    \exp(a_r) = \parens*{\frac{O(k'^{p/2})}{\alpha}}^{2/p} = \frac{O(k')}{\alpha^{2/p}} = \frac{O(k)}{\eps^{4/p}}\bracks*{(\log n)^{6/p+1} + (\log(1/\delta))^{2/p+1}}(\log(n/\delta))^{2}(\log\log n)^{4/p}.\qedhere
\]
\end{proof}

\section{Streaming and online coresets}

We present our results on streaming and online coresets for $\ell_p$ subspace approximation.

\subsection{Online coresets}
\label{sec:online-coresets}

In this section, we note that our Theorems \ref{thm:main-rls-sampling-p>2} and \ref{thm:main-rls-sampling-p<2} give the first nearly optimal online coreset construction for $\ell_p$ subspace approximation. In the online coreset model, the rows of the input matrix $\bfA$ arrive one by one. At each row arrival, we must decide whether to include the row in the coreset or not in an \emph{online fashion}, meaning we must irrevocably commit to keeping the row in the coreset or discard it permanently.

Online coresets typically depend on a quantity known as the \emph{online condition number}, which is known to be necessary in many cases \cite{CMP2016}:

\begin{Definition}[Online condition number \cite{CMP2016}]
Let $\bfA\in\mathbb R^{n\times d}$. Then, the online condition number of $\bfA$ is defined as
\[
    \kappa^\OL \coloneqq \norm{\bfA}_2 \max_{i=1}^n \norm{\bfA_{(i)}^{-}}_2,
\]
where $\bfA_{(i)}^-$ denotes the pseudoinverse of the $i\times d$ matrix formed by the first $i$ rows of $\bfA$. 
\end{Definition}

Then, the following is an immediate corollary of Theorems \ref{thm:main-rls-sampling-p>2} and \ref{thm:main-rls-sampling-p<2}.

\begin{Corollary}[Online coresets]
Let $1 \leq p < \infty$. Let $\bfA\in\mathbb R^{n\times d}$ have online condition number $\kappa^\OL$. Then, there is an online coreset algorithm which constructs a diagonal map $\bfS\in\mathbb R^{n\times n}$ satisfying Definition \ref{def:strong-coreset} with probability at least $1-\delta$, such that
\[
    \nnz(\bfS) = \begin{dcases}
        \frac{\tilde O(k^{p/2})}{\eps^{p}}(\log(n\kappa^\OL/\delta))^{O(p^2)} \\
        \frac{\tilde O(k)}{\eps^{4/p}}(\log(n\kappa^\OL/\delta))^{O(1)}
    \end{dcases}
\]
while storing at most $O(k(\log k)(\log\kappa^\OL)^2)$ additional rows in an online fashion.
\end{Corollary}
\begin{proof}
The result of \cite[Theorem 3.1]{BDMMUWZ2020} gives an online coreset algorithm for maintaining a $(1\pm\eps)$ strong coreset for $\ell_2$ subspace approximation which stores $O(\eps^{-2}k(\log k)(\log\kappa^\OL)^2)$ rows. Furthermore, given such a strong coreset with $\eps = O(1)$, it is shown in \cite[Lemma 2.11]{BDMMUWZ2020} that one can obtain scores $\tilde\bftau_i$ such that
\[
    \tilde\bftau_i \geq \bftau_i^\lambda(\bfA)
\]
for $\lambda = \norm*{\bfA-\bfA_k}_F^2/k$, and also satisfies
\[
    \sum_{i=1}^n \tilde\bftau_i \leq O(k\log\kappa^\OL).
\]
The result for $p > 2$ then follows as an immediate corollary of Theorem \ref{thm:main-rls-sampling-p>2}. For $p < 2$, we additionally need an online constant factor approximation to flatten the matrix, which is constructed in \cite{WY2023b} by obtaining online Lewis sample due to \cite{WY2023a}. The result for $p < 2$ then follows as an immediate corollary of Theorem \ref{thm:main-rls-sampling-p<2}.
\end{proof}

For integer matrices with entries bounded by $\Delta$, we may replace the dependence on the online condition number with $\Delta$, by using an analogous result of \cite{BDMMUWZ2020} for integer matrices. 

\begin{Corollary}[Online coresets -- integer matrices]
\label{cor:online-integer}
Let $1 \leq p < \infty$. Let $\bfA\in\mathbb Z^{n\times d}$ have entries bounded by $\abs*{\bfA_{i,j}}\leq\Delta$. Then, there is an online coreset algorithm which constructs a diagonal map $\bfS\in\mathbb R^{n\times n}$ satisfying Definition \ref{def:strong-coreset} with probability at least $1-\delta$, such that
\[
    \nnz(\bfS) = \begin{dcases}
        \frac{\tilde O(k^{p/2})}{\eps^{p}}(\log(n\Delta/\delta))^{O(p^2)} \\
        \frac{\tilde O(k)}{\eps^{4/p}}(\log(n\Delta/\delta))^{O(1)}
    \end{dcases}
\]
while storing at most $O(k(\log\Delta)^2)$ additional rows in an online fashion.
\end{Corollary}

\subsection{Streaming coresets}
\label{sec:streaming-coreset}

Next, we state our corollaries for constructing streaming coresets in the row arrival model of streaming, which is slightly different from the online coreset model since we are allowed to remove rows from our coreset. In the streaming model, the resource measure is typically the space complexity, and thus the input is usually assumed to be an integer matrix as a bit complexity assumption. In this setting, we combine the classic merge-and-reduce technique \cite{BDMMUWZ2020} with the technique of \cite{CWZ2023} of first applying an online coreset to obtain a result with only $\poly(\log\log(n\Delta)$ factor overhead in the coreset size.

\begin{Corollary}[Streaming coresets -- integer matrices]
Let $1 \leq p < \infty$. Let $\bfA\in\mathbb Z^{n\times d}$ have entries bounded by $\abs*{\bfA_{i,j}}\leq\Delta$. Then, there is an row arrival streaming algorithm which constructs a diagonal map $\bfS\in\mathbb R^{n\times n}$ satisfying Definition \ref{def:strong-coreset} with probability at least $1-\delta$, such that
\[
    \nnz(\bfS) = \begin{dcases}
        \frac{\tilde O(k^{p/2})}{\eps^{p}}(\log(k/\eps\delta) + \log\log(n\Delta/\delta))^{O(p^2)} \\
        \frac{\tilde O(k)}{\eps^{4/p}}(\log(k/\eps\delta) + \log(n\Delta/\delta))^{O(1)}
    \end{dcases}
\]
while storing at most $O(k(\log\Delta)^2)$ additional rows in an online fashion.
\end{Corollary}
\begin{proof}
We may assume without loss of generality that the stream length is at most $m = \poly(k, \eps^{-1}, \log(n\Delta/\delta))$ by first applying Corollary \ref{cor:online-integer}. We then apply the merge-and-reduce technique, which results in a coreset with size where the $\eps$ dependence is replaced by $\eps' = \eps/\log m$. This results in the claimed bounds.
\end{proof}

\section*{Acknowledgements}

David P. Woodruff and Taisuke Yasuda were supported in part by NSF CCF-2335411 and a Simons Investigator Award.

\bibliographystyle{alpha}
\bibliography{citations}

\appendix

\section{\texorpdfstring{$\ell_p$}{lp} affine embeddings via root leverage score sampling}

Following the techniques of \cite{WY2023c}, we obtain a root leverage score sample guarantee for $\ell_p$ affine embeddings, for both $p<2$ and $p>2$. 

Our main result of this section is the following theorem, which shows that root leverage score sampling yields $\ell_p$ affine embeddings. 

\begin{Theorem}[Root leverage score sampling]
\label{thm:root-leverage-score-sampling-affine}
Let $\bfA\in\mathbb R^{n\times d}$ and $\bfb\in\mathbb R^n$. Let $1\leq p < \infty$. Let $R\geq\norm{\bfb}_p$. Suppose that 
\[
    \frac{\abs{\bfb(i)}^p}{R^p} \leq \begin{dcases}
        \min\braces*{1, n^{p/2-1}\bftau_i(\bfA)^{p/2}} & p > 2 \\
        \bftau_i(\bfA)^{p/2} & p < 2
    \end{dcases}
\]
for every $i\in[n]$. Let $\alpha = \Theta(\eps^2) / ((\log n)^3 + \log(1/\delta))$ and let
\[
    q_i \geq \begin{dcases}
        \min\braces*{1, n^{p/2-1}\bftau_i(\bfA)^{p/2}/\alpha} & p > 2 \\
        \min\braces*{1, \bftau_i(\bfA)^{p/2}/\alpha} & p < 2
    \end{dcases}
\]
Let $\bfS$ be the $\ell_p$ sampling matrix (Definition \ref{def:sampling-matrix}) with sampling probabilities $\{q_i\}_{i=1}^n$. Then, with probability at least $1-\delta$, for every $\bfx\in\mathbb R^d$,
\[
    \norm*{\bfS(\bfA\bfx+\bfb)}_p^p = (1\pm \eps) \norm*{\bfA\bfx+\bfb}_p^p \pm \eps R^p
\]
\end{Theorem}

\subsection{Reduction to a small number of scales}

Our first task is to reduce the proof of Theorem \ref{thm:root-leverage-score-sampling-affine} to showing a similar theorem when $\bfA\bfx$ is restricted to a certain scale, for a small number of scales. This is shown in the following lemma:

\begin{Lemma}
\label{lem:reduction-to-scales}
Let $\bfA\in\mathbb R^{n\times d}$ and $\bfb\in\mathbb R^n$. Let $1\leq p < \infty$. Let $R_0 \geq \norm*{\bfb}_p$ and $0 < \eps < 1/2$. Suppose that
\[
    \sup_{\norm*{\bfA\bfx}_p = 1} \abs*{\norm*{\bfS\bfA\bfx}_p^p - \norm*{\bfA\bfx}_p^p} \leq \eps
\]
and that
\[
    \sup_{\norm*{\bfA\bfx}_p \leq R_i} \abs*{\norm*{\bfS(\bfA\bfx+\bfb)}_p^p - \norm*{\bfA\bfx+\bfb}_p^p} \leq \eps R_i^p
\]
holds for each $R_i = 2^i \cdot R_0$, $i\in[I]$, where $I = O(\log \eps^{-1})$. Then,
\[
    \norm*{\bfS(\bfA\bfx+\bfb)}_p^p = (1\pm 2\cdot 4^p\eps) \norm*{\bfA\bfx+\bfb}_p^p \pm 2^p \eps R_0^p
\]
for every $\bfx\in\mathbb R^d$.
\end{Lemma}
\begin{proof}
First note that if $\norm*{\bfA\bfx}_p \leq R_1$, then we immediately have
\[
    \norm*{\bfS(\bfA\bfx+\bfb)}_p^p = \norm*{\bfA\bfx+\bfb}_p^p \pm \eps R_1^p = \norm*{\bfA\bfx+\bfb}_p^p \pm 2^p \eps R_0^p.
\]
Next, suppose that $\norm*{\bfA\bfx}_p \geq R_0 / \eps$. Note that
\[
    \norm*{\bfA\bfx+\bfb}_p = \norm*{\bfA\bfx}_p \pm \norm*{\bfb}_p = (1\pm\eps)\norm*{\bfA\bfx}_p
\]
and similarly,
\[
    \norm*{\bfS(\bfA\bfx+\bfb)}_p = \norm*{\bfS\bfA\bfx}_p \pm \norm*{\bfS\bfb}_p = (1\pm 4\eps)\norm*{\bfS\bfA\bfx}_p
\]
Thus,
\[
    \norm*{\bfS(\bfA\bfx+\bfb)}_p = (1\pm 4\eps)(1\pm\eps)\norm*{\bfA\bfx+\bfb}_p = (1\pm 7\eps)\norm*{\bfA\bfx+\bfb}_p.
\]
Finally, we handle the intermediate scales between $R_0$ and $R_0/\eps$. Consider $\bfx$ such that $R_i \leq \norm*{\bfA\bfx}_p < 2\cdot R_i$. Note then that
\[
    \norm*{\bfA\bfx+\bfb}_p \geq \norm*{\bfA\bfx}_p -\norm*{\bfb}_p \geq \norm*{\bfA\bfx}_p / 2 \geq R_i / 2
\]
so
\[
    \norm*{\bfS(\bfA\bfx+\bfb)}_p^p = \norm*{\bfA\bfx+\bfb}_p^p \pm \eps \cdot (2R_i)^p = \norm*{\bfA\bfx+\bfb}_p^p \pm \eps \cdot (4\norm*{\bfA\bfx+\bfb}_p)^p.
\]
This covers all cases.
\end{proof}

\subsection{Reduction to a Gaussian process with flat sensitivities}

We now work towards bounding a quantity as the one used in Lemma \ref{lem:reduction-to-scales}. The following lemma is a standard reduction in chaining arguments (see, e.g., Lemma D.1 of \cite{WY2023c}, also \cite{CP2015, CD2021}):

\begin{Lemma}[Reduction to Gaussian processes]
\label{lem:gp-reduction}
Let $\bfA\in\mathbb R^{n\times d}$ and $\bfb\in\mathbb R^n$. Let $1\leq p < \infty$. Let $R\geq\norm{\bfb}_p$. Let $\bfS$ be a random $\ell_p$ sampling matrix (Definition \ref{def:sampling-matrix}). Then,
\[
    \E_{\bfS}\sup_{\norm*{\bfA\bfx}_p \leq R} \abs*{\norm*{\bfS(\bfA\bfx+\bfb)}_p^p - \norm*{\bfA\bfx+\bfb}_p^p}^l \leq (2\pi)^{l/2} \E_{\bfS}\E_{\bfg\sim\mathcal N(0,\bfI_n)}\sup_{\norm*{\bfA\bfx}_p \leq R} \abs*{\sum_{i\in T} \bfg_i \abs*{[\bfS(\bfA\bfx+\bfb)](i)}^p}^l,
\]
where $T\subseteq[n]$ is the set of rows with sampling probability $q_i < 1$.
\end{Lemma}

We will further reduce the problem to a similar problem for an instance with ``flat sensitivities'', following \cite{CP2015, WY2023c}. We will use the following trivial generalization of the result of \cite[Lemma D.5]{WY2023c}.

\begin{Lemma}[Flattening all sensitivities]
\label{lem:flatten-all}
Let $1 \leq p < \infty$ and $\bfA\in\mathbb R^{n\times d}$ and $\bfb\in\mathbb R^n$. Let $0<\alpha<1$. Then, there exists $\bfA'\in\mathbb R^{m\times d}$ and $\bfb'\in\mathbb R^m$ for $m = O(n\alpha^{-1})$ such that
\[
    \bfsigma_i^q(\bfA') \leq \alpha, \qquad \bfsigma_i^q(\bfb') = \frac{\abs{\bfb(i)}^q}{\norm*{\bfb}_q^q}\leq \alpha
\]
for every $i\in[m]$ and $1\leq q < \infty$. Furthermore, for any $1 \leq q < \infty$ and $\bfx\in\mathbb R^d$, we have that $\norm*{\bfA'\bfx+\bfb'}_q = \Theta(\alpha^{1/p-1/q})\norm*{\bfA\bfx+\bfb}_q$.
\end{Lemma}

Using Lemma \ref{lem:flatten-all}, we construct the following new instance with bounded $\ell_2$ and $\ell_p$ sensitivities:

\begin{Lemma}[Flattened instance]
\label{lem:flattened-instance}
Let $\bfA\in\mathbb R^{n\times d}$ and $\bfb\in\mathbb R^{n\times d}$. Let $R\geq \norm*{\bfb}_p$. Suppose that 
\[
    \frac{\abs{\bfb(i)}^p}{R^p} \leq \begin{dcases}
        \min\braces*{1, n^{p/2-1}\bftau_i(\bfA)^{p/2}} & p > 2 \\
        \bftau_i(\bfA)^{p/2} & p < 2
    \end{dcases}
\]
for every $i\in[n]$. Let $0<\alpha<1$ and let
\[
    q_i \geq \begin{dcases}
        \min\braces*{1, n^{p/2-1}\bftau_i(\bfA)^{p/2}/\alpha} & p > 2 \\
        \min\braces*{1, \bftau_i(\bfA)^{p/2}/\alpha} & p < 2
    \end{dcases}
\]
Let $T\subseteq[n]$ be the set of rows $i\in[n]$ with $q_i < 1$. Let $\bfS$ be a diagonal matrix with $\bfS_{i,i} \leq 1/q_i^{1/p}$. Then, there is $\bfA''\in\mathbb R^{m\times d}$ and $\bfb''\in\mathbb R^m$ for $m = O(n/\alpha)$ such that
\begin{itemize}
    \item $\bftau_i(\bfA'') \leq O(\alpha)$ for $p<2$ and $\bftau_i(\bfA'') \leq O(\alpha) / n^{1-2/p}$ for $p>2$, for every $i\in[m]$
    \item $\bfsigma_i^p(\bfA'') \leq O(\alpha)$ and $\abs{\bfb''(i)}^p / R^p\leq O(\alpha)$ for every $i\in[m]$
    \item $\norm*{\bfA''\bfx+\bfb''}_p^p = \norm*{\bfA\bfx+\bfb}_p^p + \norm*{\bfS\vert_T(\bfA\bfx+\bfb)}_p^p$ for every $\bfx\in\mathbb R^d$
    \item $\norm*{\bfA''\bfx}_p^p = \norm*{\bfA\bfx}_p^p + \norm*{\bfS\vert_T\bfA\bfx}_p^p$ for every $\bfx\in\mathbb R^d$
\end{itemize}
\end{Lemma}

\begin{proof}
Let $\bfA'\in\mathbb R^{m\times d}$ and $\bfb'\in\mathbb R^m$ be the flattened instances given by Lemma \ref{lem:flatten-all}, where $m = O(n/\alpha)$. Now let
\[
    \bfA'' \coloneqq \begin{pmatrix}
    \bfA' \\ \bfS\vert_T\bfA
    \end{pmatrix}, \qquad \bfb'' \coloneqq \begin{pmatrix}
    \bfb' \\ \bfS\vert_T\bfb
    \end{pmatrix}
\]
be the $(m + n_\bfS) \times d$ matrix and $(m+n_\bfS)$-dimensional vector formed by the vertical concatenation of $\bfA'$ and $\bfb'$ with $\bfS\bfA$ and $\bfS\bfb$, where $n_\bfS$ is the number of rows sampled by $\bfS$.

We now show how to bound the sensitivities of $\bfA''$ and $\bfb''$.

For any row $i$ corresponding to a row of $\bfA'$, the $\ell_2$ sensitivities are already bounded by $\alpha$, and furthermore, $\ell_2$ sensitivities can clearly only decrease with row additions. For any row $i$ corresponding to a row of $\bfS\bfA$ that is sampled with probability $q_i < 1$, we have that
\[
    \frac{\abs*{[\bfS\bfA\bfx](i)}^2}{\norm*{\bfA''\bfx}_2^2} \leq \frac{\abs*{[\bfS\bfA\bfx](i)}^2}{\norm*{\bfA'\bfx}_2^2} = \frac{\abs*{[\bfS\bfA\bfx](i)}^2}{\Theta(\alpha^{2/p-1})\norm*{\bfA\bfx}_2^2} \leq \frac1{q_i^{2/p}}\frac{\abs*{[\bfA\bfx](i)}^2}{\Theta(\alpha^{2/p-1})\norm*{\bfA\bfx}_2^2} \leq \frac{\bftau_i(\bfA)}{\Theta(\alpha^{2/p-1}) q_i^{2/p}} = O(\alpha).
\]
In fact, for $p>2$, we have the stronger bound of
\[
    \frac{\bftau_i(\bfA)}{\Theta(\alpha^{2/p-1}) q_i^{2/p}} \leq \frac{O(\alpha)}{n^{1-2/p}}.
\]
Thus, we have that $\bftau_i(\bfA'') = \bfsigma_i^2(\bfA'') \leq O(\alpha)$ for every row $i$ of $\bfA''$.

For $p<2$, the max sensitivity is bounded by $O(\alpha)$ by the monotonicity of max sensitivities. For $p>2$, we have by reverse monotonicity of max sensitivities that
\[
    \bfsigma_i^p(\bfA) \leq n^{p/2-1}\bftau_i(\bfA)
\]
so for any row $i$ corresponding to a row of $\bfS\bfA$ sampled with probability $q_i < 1$, we have that
\[
    \frac{\abs*{[\bfS\bfA\bfx](i)}^p}{\norm*{\bfA''\bfx}_p^p} \leq \frac{\abs*{[\bfS\bfA\bfx](i)}^p}{\norm*{\bfA'\bfx}_p^p} = \frac{\abs*{[\bfS\bfA\bfx](i)}^p}{\norm*{\bfA\bfx}_p^p} \leq \frac1{q_i}\frac{\abs*{[\bfA\bfx](i)}^p}{\norm*{\bfA\bfx}_p^p} \leq \frac{\bfsigma_i^p(\bfA)}{q_i} \leq O(\alpha).
\]
By similar reasoning, we have that
\[
    \frac{\abs{[\bfS\bfb](i)}^p}{R^p} \leq \frac1{q_i}\frac{\abs{\bfb(i)}^p}{R^p} \leq O(\alpha).\qedhere
\]
\end{proof}

The next lemma shows that in order to bound the Gaussian process in Lemma \ref{lem:gp-reduction}, it suffices to bound a similar Gaussian process for $\bfA''$ and $\bfb''$. 

\begin{Lemma}[Reduction to flattened instance]
\label{lem:flat-gp-reduction}
Let $\bfS$, $\bfA,\bfb$, and $\bfA'',\bfb''$ be as given in Lemma \ref{lem:flattened-instance}. Let $\delta, \eps, l$ be such that $\delta\eps^l \leq 1/2$. Furthermore, suppose that
\[
    \E_{\bfS}\E_{\bfg\sim\mathcal N(0,\bfI_n)}\sup_{\norm*{\bfA''\bfx}_p \leq R} \abs*{\sum_{i\in T} \bfg_i \abs*{[\bfA''\bfx+\bfb](i)}^p}^l \leq \delta \eps^l (R^p)^l
\]
for every $R\geq \norm{\bfb}_p$. Then,
\[
    \E_{\bfS}\sup_{\norm*{\bfA\bfx}_p \leq R} \abs*{\norm*{\bfS(\bfA\bfx+\bfb)}_p^p - \norm*{\bfA\bfx+\bfb}_p^p}^l \leq 2\delta (2^{3p} \eps R^p)^l
\]
for every $R\geq\norm*{\bfb}_p$.
\end{Lemma}
\begin{proof}
Fix an outcome of $\bfS$ and let
\[
    F_{\bfS, R} = \sup_{\norm*{\bfA\bfx}_p \leq R} \abs*{\norm*{\bfS(\bfA\bfx+\bfb)}_p^p - \norm*{\bfA\bfx+\bfb}_p^p}.
\]
Note then that for any $\norm*{\bfA\bfx}_p\leq R$, we have
\begin{align*}
    \norm*{\bfA''\bfx}_p &\leq \norm*{\bfA\bfx}_p + \norm*{\bfS(\bfA\bfx+\bfb)}_p + \norm*{\bfS\bfb}_p \\
    &\leq R + \parens*{\norm*{\bfA\bfx+\bfb}_p^p + \abs*{\norm*{\bfS(\bfA\bfx+\bfb)}_p^p - \norm*{\bfA\bfx+\bfb}_p^p}}^{1/p} + \parens*{\norm*{\bfb}_p^p + \abs*{\norm*{\bfS\bfb}_p^p - \norm*{\bfb}_p^p}}^{1/p} \\
    &\leq R + \parens*{\norm*{\bfA\bfx+\bfb}_p^p + F_{\bfS,R}}^{1/p} + \parens*{\norm*{\bfb}_p^p + F_{\bfS,R}}^{1/p} \\
    &\leq R + \norm*{\bfA\bfx+\bfb}_p + F_{\bfS,R}^{1/p} + \norm*{\bfb}_p + F_{\bfS,R}^{1/p} \\
    &\leq 4R + 2F_{\bfS,R}^{1/p}.
\end{align*}
Thus,
\begin{align*}
    \E_{\bfS}F_{\bfS,R}^l &\leq \E_{\bfS}\E_{\bfg\sim\mathcal N(0,\bfI_n)}\sup_{\norm*{\bfA\bfx}_p \leq R} \abs*{\sum_{i\in T} \bfg_i \abs*{[\bfS(\bfA\bfx+\bfb)](i)}^p}^l && \text{Lemma \ref{lem:gp-reduction}} \\
    &\leq \E_{\bfS}\E_{\bfg\sim\mathcal N(0,\bfI_n)}\sup_{\norm*{\bfA\bfx}_p \leq R} \abs*{\sum_{i} \bfg_i \abs*{[\bfA''\bfx+\bfb''](i)}^p}^l \\
    &\leq \E_{\bfS}\E_{\bfg\sim\mathcal N(0,\bfI_n)}\sup_{\norm*{\bfA''\bfx}_p \leq 4R + 2F_{\bfS,R}^{1/p}} \abs*{\sum_{i} \bfg_i \abs*{[\bfA''\bfx+\bfb''](i)}^p}^l \\
    &\leq \E_{\bfS}\delta \eps^l ((4R + 2F_{\bfS,R}^{1/p})^p)^l && \text{by hypothesis} \\
    &\leq \E_{\bfS}\delta (2^{2p}\eps)^l (((2R)^p)^l + F_{\bfS,R}^l) \\
    &= \delta (2^{2p}\eps)^l \bracks*{((2R)^p)^l + \E_{\bfS} F_{\bfS,R}^l}
\end{align*}
so rearranging gives
\[
    \frac{\E_{\bfS} F_{\bfS,R}^l}{(2^p R^p)^l + \E_{\bfS} F_{\bfS,R}^l} \leq \delta (2^{2p}\eps)^l.
\]
In turn, this implies that
\[
    \E_{\bfS} F_{\bfS,R}^l \leq \frac{\delta (2^{2p}\eps)^l (2^p R^p)^l}{1 - \delta (2^{2p}\eps)^l} \leq 2\delta (2^{3p} \eps R^p)^l.\qedhere
\]
\end{proof}

\subsection{Bounds on the Gaussian process}

In this section, we present results from \cite{WY2023c} (which in turn are based on \cite{BLM1989, LT1991}) which will allow us to bound a Gaussian process of the form of Lemma \ref{lem:flat-gp-reduction}.

The following is a straightforward generalization of Lemma C.3 of \cite{WY2023c}.

\begin{Lemma}
\label{lem:dx-bound}
Let $1\leq p < \infty$ and let $\bfA\in\mathbb R^{n\times d}$ and $\bfb\in\mathbb R^n$. Let $R \geq \norm*{\bfb}_p$. Define the pseudo-metric
\[
    d_X(\bfy,\bfy') \coloneqq \parens*{\E_{\bfg\sim\mathcal N(0,\bfI_n)}\abs*{\sum_{i=1}^n  \bfg_i \abs{\bfy(i)}^p - \sum_{i=1}^n  \bfg_i \abs{\bfy'(i)}^p}^2}^{1/2}
\]
Let $\sigma \geq \max_{i\in S}^n \bfsigma_i^p(\bfA) + \abs{\bfb(i)}^p/R^p$. Then, for any $\bfy = \bfA\bfx+\bfb$ and $\bfy' = \bfA\bfx'+\bfb$ with $\norm*{\bfA\bfx}_p,\norm*{\bfA\bfx'}_p \leq R$,
\[
    d_X(\bfy,\bfy') \leq \begin{dcases}
        O(1) \norm*{\bfA(\bfx-\bfx')}_\infty^{p/2}R^{p/2} & p < 2 \\
        O(1)\sigma^{1/2-1/p}\cdot \norm*{\bfA(\bfx-\bfx')}_\infty R^{p-1}  & p > 2
    \end{dcases}
\]
\end{Lemma}

With Lemma \ref{lem:dx-bound} in hand, the following result follows immediately from the proofs in \cite{WY2023c}:

\begin{Theorem}
\label{thm:dudley-bound}
Let $1 \leq p < \infty$ with $p\neq 2$ be fixed and let $\bfA\in\mathbb R^{n\times d}$ and $\bfb\in\mathbb R^n$. Let $R\geq\norm{\bfb}_p$. Let $\tau \geq \max_{i=1}^n\bftau_i(\bfA)$ and let $\sigma \geq \max_{i=1}^n \bfsigma_i^p(\bfA)$. Define
\[
    \mathcal E \coloneqq \begin{dcases}
        \tau^{1/2} (\sigma n)^{1/2-1/p} \cdot (\log n)^{3/2} & p > 2 \\
        \tau^{1/2} \cdot \parens*{ \log n}^{3/2} & p < 2 \\
    \end{dcases}.
\]
Then,
\[
    \E_{\bfg\sim\mathcal N(0,\bfI_n)}\sup_{\norm*{\bfA\bfx}_p \leq R}\abs*{\sum_{i=1}^n  \bfg_i \abs{[\bfA\bfx+\bfb](i)}^p}^l \leq \bracks*{(2\mathcal E)^l (\mathcal E / \sigma^{1/2}) + O(\sqrt l \sigma^{1/2})^l} (R^p)^l
\]
\end{Theorem}

\subsection{Proof of main sampling theorems}

We now prove Theorem \ref{thm:root-leverage-score-sampling-affine} by combining the previous results of this section.

\begin{proof}[Proof of Theorem \ref{thm:root-leverage-score-sampling-affine}]
By Lemma \ref{lem:reduction-to-scales}, it suffices to show that
\begin{equation}
\label{eq:scale-guarantee}
    \sup_{\norm*{\bfA\bfx}_p \leq R_i} \abs*{\norm*{\bfS(\bfA\bfx+\bfb)}_p^p - \norm*{\bfA\bfx+\bfb}_p^p} \leq \eps R_i^p
\end{equation}
for $R_i = 2^i R$ for $i\in[I]$, $I = O(\log\eps^{-1})$. The corresponding statement for bounding
\[
    \sup_{\norm*{\bfA\bfx}_p = 1} \abs*{\norm*{\bfS\bfA\bfx}_p^p - \norm*{\bfA\bfx}_p^p} \leq \eps 
\]
will follow from the exact same analysis by setting $\bfb = 0$ and $R = 0$. 

In order to obtain \eqref{eq:scale-guarantee} for a single scale with high probability, we will bound the $l$th moment for a large even power $l$. We will bound this quantity by passing to a Gaussian process bound in Lemma \ref{lem:flat-gp-reduction}, which we can in turn bound using Theorem \ref{thm:dudley-bound}. Note that by our construction of the flattened instance $\bfA''$ and $\bfb''$ in Lemmas \ref{lem:flattened-instance} and \ref{lem:flat-gp-reduction}, we have $\tau = O(\alpha)$ and $\sigma = O(\alpha)$ for $p < 2$ and $\tau = O(\alpha) / n^{1-2/p}$ and $\sigma = O(\alpha)$ for $p>2$, so we can bound the $\mathcal E$ parameter in Theorem \ref{thm:dudley-bound} by
\[
    \mathcal E\leq \begin{cases}
    O(\alpha^{1-1/p})(\log n)^{3/2} & p > 2 \\
    O(\alpha^{1/2})(\log n)^{3/2} & p < 2
    \end{cases}
\]
In turn, the bound on the Gaussian process in Theorem \ref{thm:dudley-bound} is $\delta \eps^l (R_i^p)^l / (I + 1)$ by our choice of $\alpha$, for 
\[
    l = O\parens*{\log\log n + \log\log\frac1\eps + \log\frac1\delta}.
\]
That is, we have shown that
\[
    \E_{\bfS}\sup_{\norm*{\bfA\bfx}_p \leq R_i} \abs*{\norm*{\bfS(\bfA\bfx+\bfb)}_p^p - \norm*{\bfA\bfx+\bfb}_p^p}^l \leq \frac{\delta}{I+1} \eps^l (R_i^p)^l.
\]
Then by Markov's inequality, we have that
\[
    \Pr_\bfS\braces*{\sup_{\norm*{\bfA\bfx}_p \leq R_i} \abs*{\norm*{\bfS(\bfA\bfx+\bfb)}_p^p - \norm*{\bfA\bfx+\bfb}_p^p} \leq \eps R_i^p} \geq 1 - \frac{\delta}{I+1}.
\]
Now by a union bound, this is simultaneously true for all $i\in [I]$ as well as for $\bfb = 0$ and $R = 0$ by a union bound, all with probability at least $1 - \delta$. In turn, we have that
\[
    \norm*{\bfS(\bfA\bfx+\bfb)}_p^p = (1\pm 2\cdot 4^p\eps) \norm*{\bfA\bfx+\bfb}_p^p \pm 2^p \eps R_0^p
\]
by Lemma \ref{lem:reduction-to-scales}. We conclude with the desired conclusion by rescaling $\eps$ and $\delta$ up to constant factors.
\end{proof}

\section{Sharper bounds for representative subspaces}
\label{sec:sharper-sohler-woodruff}

We provide sharper bounds for the result of \cite[Theorem 10]{SW2018}.

\begin{Theorem}[Representative subspace theorem]
\label{thm:representative-subspace}
Let $1 \leq p < \infty$. Suppose that an $s$-dimensional subspace $S$ satisfies
\[
    \norm*{\bfA(\bfP_S-\bfP_{S\cup F})}_{p,2}^p \leq \eps^{p}\cdot \OPT
\]
for every $F\in\mathcal F_k$. Then if $\bfP_S$ is the projection matrix onto $S$ and $\bfb_S \in\mathbb R^n$ is the vector defined by
\[
    \bfb_S(i) \coloneqq \norm*{\bfa_i^\top(\bfI - \bfP_S)}_2,
\]
then
\begin{equation}\label{eq:sw18-guarantee}
    \mbox{for all $F\in\mathcal F_k$,} \qquad \norm*{\bfA(\bfI-\bfP_F)}_{p,2}^p = (1\pm\eps)\norm*{[\bfA\bfP_S(\bfI-\bfP_F),~\bfb_S]}_{p,2}^p,
\end{equation}
where $[\bfA\bfP_S(\bfI-\bfP_F),~\bfb_S]$ denotes the $n\times (d+1)$ concatenation of $\bfA\bfP_S(\bfI-\bfP_F)\in\mathbb R^{n\times d}$ and $\bfb_S\in\mathbb R$. Furthermore, such a subspace $S$ exists for
\[
    s = \frac{O(k)}{\eps^{\max\{2, p\}}}
\]
such that $\norm*{\bfb_S}_p^p \leq \OPT$.
\end{Theorem}

\subsection{Sharper scalar inequalities}

The following result simplifies and sharpens \cite[Claim 2]{SW2018}.

\begin{Lemma}
\label{lem:p-diff}
Let $u, v, w\geq 0$ satisfy $u^2 = v^2 - w^2$. Then,
\[
    u^p \leq \begin{cases}
        \min\{\eps v^p, 2^{p-1}\eps^{1-2/p} (v^p - w^p)\} & 1 \leq p \leq 2 \\
        v^p - w^p & 2 \leq p < \infty
    \end{cases}
\]
\end{Lemma}
\begin{proof}
The second inequality follows from the subadditivity of $(\cdot)^{p/2}$ \cite{SW2018} so it remains to show the first. We may assume that $v = 1$ by scaling. We also reparameterize $w = 1 - x$ for some $0\leq x\leq 1$. Then,
\[
    u^p = (1 - (1-x)^2)^{p/2} \leq (2x)^{p/2}
\]
and
\[
    \frac{u^p}{v^p - w^p} = \frac{u^p}{1 - (1-x)^p} \leq \frac{(2x)^{p/2}}{x} = 2^{p/2} x^{p/2-1}
\]
Thus, if $x \leq \eps^{2/p}/2$, then $u^p \leq \eps$, and if $x\geq \eps^{2/p}/2$, then $u^p \leq 2^{p-1}\eps^{1-2/p}(v^p - w^p)$.
\end{proof}

The following result sharpens \cite[Claim 5]{SW2018}.

\begin{Lemma}
\label{lem:eps-tri-ineq}
Let $u, v\geq 0$ and $1\leq p < \infty$. Then,
\[
    (u+v)^p \leq (1+\eps) u^p + \frac{(2p)^p}{\eps^{p-1}} v^p
\]
\end{Lemma}
\begin{proof}
We may assume that $u = 1$ by scaling. If $v\geq 1$, then $(1+v)^p \leq 2^p v^p$ so assume that $v\leq 1$. Then, $(1+v)^p \leq 1 + 2pv$ so if $2pv \leq \eps$, then $(1+v)^p \leq \eps$, while if $2pv \geq \eps$, then
\[
    (1+v)^p \leq 1 + 2pv = 1 + \frac{2p}{v^{p-1}}v^p \leq 1 + \frac{(2p)^p}{\eps^{p-1}} v^p.
\]
\end{proof}

The following result generalizes \cite[Lemma 4]{SW2018} to $p > 1$.

\begin{Lemma}
\label{lem:root-exch}
Let $a, b, f, g\geq 0$ and $1\leq p < \infty$. Then,
\[
    \abs{(a^2+b^2)^{p/2} - (f^2+g^2)^{p/2}} \leq \frac{(4p)^p}{2\eps^{p-1}}\parens*{\abs{a-f}^p + \abs{b-g}^p} + \eps \parens*{(a^2+b^2)^{p/2} + (f^2+g^2)^{p/2}}.
\]
\end{Lemma}
\begin{proof}
By Lemma \ref{lem:eps-tri-ineq}, we have that
\[
    \norm{(a,b)}_2^p \leq (\norm{(a-f,b-g)}_2 + \norm{(f,g)}_2)^p \leq (1+\eps)\norm{(f,g)}_2^p + \frac{(2p)^p}{\eps^{p-1}}\norm{(a-f,b-g)}_2^p
\]
and similarly
\[
    \norm{(f,g)}_2^p \leq (\norm{(a-f,b-g)}_2 + \norm{(a,b)}_2)^p \leq (1+\eps)\norm{(a,b)}_2^p + \frac{(2p)^p}{\eps^{p-1}}\norm{(a-f,b-g)}_2^p.
\]
Thus,
\[
    \abs{\norm{(a,b)}_2^p - \norm{(f,g)}_2^p} \leq \frac{(2p)^p}{\eps^{p-1}}\norm{(a-f,b-g)}_2^p + \eps (\norm{(a,b)}_2^p + \norm{(f,g)}_2^p).
\]
Finally, we bound
\[
    \norm{(a-f,b-g)}_2^p \leq \norm{(a-f,b-g)}_1^p \leq 2^{p-1}(\abs{a-f}^p + \abs{b-g}^p).
\]
\end{proof}

\subsection{Representative subspace theorem}

The first lemma shows that if $\norm{\bfa^\top(\bfP_S - \bfP_{S\cup F})}_2$ is small, then the projection of a vector $\bfa$ onto $S\cup F$ is close to its projection onto $S$, and the projection of $\bfa^\top \bfP_{S\cup F}$ onto $F$ is close to the projection of $\bfa^\top \bfP_S$ onto $F$. 

\begin{Lemma}
\label{lem:close-proj}
Let $S, F\subseteq\mathbb R^d$ be subspaces and let $\bfa\in\mathbb R^d$ be a vector. Then,
\begin{itemize}
\item $\norm{\bfa^\top(\bfI - \bfP_{S\cup F})}_2 = \norm{\bfa^\top(\bfI - \bfP_S)}_2 \pm \norm{\bfa^\top(\bfP_S - \bfP_{S\cup F})}_2$
\item $\norm{\bfa^\top(\bfP_{S\cup F} - \bfP_{F})}_2 = \norm{\bfa^\top \bfP_S(\bfI - \bfP_F)}_2 \pm \norm{\bfa^\top(\bfP_S - \bfP_{S\cup F})}_2$
\end{itemize}
\end{Lemma}
\begin{proof}
These are proven in \cite{SW2018}. We reproduce a proof for the reader's convenience. The first inequality is just the triangle inequality, so it remains to show the latter. One direction of the inequality follows by
\begin{align*}
    \norm{\bfa^\top(\bfP_{S\cup F} - \bfP_F)}_2 &= \min_{\bfx\in F}\norm{\bfa^\top\bfP_{S\cup F} - \bfx}_2 \\
    &\leq \norm{\bfa^\top\bfP_{S\cup F} - \bfa^\top\bfP_S\bfP_F}_2 \leq \norm{\bfa^\top(\bfP_{S\cup F} - \bfP_S)}_2 + \norm{\bfa^\top\bfP_S(\bfI - \bfP_F)}_2
\end{align*}
and the other by
\begin{align*}
    \norm{\bfa^\top\bfP_{S}(\bfI - \bfP_F)}_2 &= \min_{\bfx\in F}\norm{\bfa^\top\bfP_{S} - \bfx}_2 \\
    &\leq \norm{\bfa^\top\bfP_{S} - \bfa^\top\bfP_F}_2 \leq \norm{\bfa^\top(\bfP_{S} - \bfP_{S\cup F})}_2 + \norm{\bfa^\top(\bfP_{S\cup F} - \bfP_F)}_2
\end{align*}
\end{proof}

We may combine Lemma \ref{lem:close-proj} with Lemma \ref{lem:root-exch} to show the following, which states that the projection cost of $\bfa$ onto $F$ is approximately the sum of the cost of projecting onto $S$, and then projecting onto $F$.

\begin{Lemma}
\label{lem:proj-bound}
Let $S, F\subseteq\mathbb R^d$ be subspaces and let $\bfa\in\mathbb R^d$ be a vector. Then,
\begin{align*}
    \abs*{\norm{\bfa^\top(\bfI-\bfP_F)}_2^p - (\norm{\bfa^\top(\bfI-\bfP_{S})}_2^2 + \norm{\bfa^\top\bfP_{S}(\bfI-\bfP_F)}_2^2)^{p/2}} &\leq \parens*{\frac{(4p)^p}{\eps^{p-1}} + 2^{p-1}\eps}\norm{\bfa^\top(\bfP_S-\bfP_{S\cup F})}_2^p \\
    &\hspace{5em}+ (2^{p-1}+1)\eps\norm{\bfa^\top(\bfI-\bfP_F)}_2^p
\end{align*}
\end{Lemma}
\begin{proof}
Note that by orthogonality,
\[
    \norm{\bfa^\top(\bfI-\bfP_F)}_2^2 = \norm{\bfa^\top(\bfI-\bfP_{S\cup F})}_2^2 + \norm{\bfa^\top(\bfP_{S\cup F}-\bfP_F)}_2^2.
\]
Then, we apply Lemma \ref{lem:root-exch} with $a = \norm{\bfa^\top(\bfI-\bfP_{S\cup F})}_2$, $b = \norm{\bfa^\top(\bfP_{S\cup F}-\bfP_F)}_2$, $f = \norm{\bfa^\top(\bfI-\bfP_{S})}_2$, and $g = \norm{\bfa^\top\bfP_{S}(\bfI-\bfP_F)}_2$ as well as the bound
\[
    \abs{a-b}, \abs{f-g} \leq \norm{\bfa^\top(\bfP_S-\bfP_{S\cup F})}_2
\]
from Lemma \ref{lem:close-proj} to see that
\begin{align*}
    \abs{\norm{(a,b)}_2^2 - \norm{(f,g)}_2^2}^{p/2} &= \abs*{\norm{\bfa^\top(\bfI-\bfP_F)}_2^p - (\norm{\bfa^\top(\bfI-\bfP_{S})}_2^2 + \norm{\bfa^\top\bfP_{S}(\bfI-\bfP_F)}_2^2)^{p/2}} \\
    &\leq \frac{(4p)^p}{2\eps^{p-1}}(\abs{a-f}^p + \abs{b-g}^p) + \eps(\norm{(a,b)}_2^p + \norm{(f,g)}_2^p) \\
    &\leq \frac{(4p)^p}{\eps^{p-1}}\norm{\bfa^\top(\bfP_S-\bfP_{S\cup F})}_2^p  \\
    &\hspace{5em}+\eps(\norm{\bfa^\top(\bfI-\bfP_F)}_2^p +(\norm{\bfa^\top(\bfI-\bfP_{S})}_2^2 + \norm{\bfa^\top\bfP_{S}(\bfI-\bfP_F)}_2^2)^{p/2}).
\end{align*}
Note that
\begin{align*}
    (\norm{\bfa^\top(\bfI-\bfP_{S})}_2^2 + \norm{\bfa^\top\bfP_{S}(\bfI-\bfP_F)}_2^2)^{p/2} &\leq 2^{p-1}(\norm{\bfa^\top(\bfI-\bfP_{S\cup F})}_2^2 + \norm{\bfa^\top(\bfP_{S\cup F}-\bfP_F)}_2^2)^{p/2} \\
    &\hspace{5em}+ 2^{p-1}\norm{\bfa^\top(\bfP_S-\bfP_{S\cup F})}_2^p \\
    &= 2^{p-1}\norm{\bfa^\top(\bfI-\bfP_F)}_2^p + 2^{p-1}\norm{\bfa^\top(\bfP_S-\bfP_{S\cup F})}_2^p
\end{align*}
so combining the bounds gives the claimed result.
\end{proof}

It remains to construct a subspace $S$ such that $\norm{\bfA(\bfP_S-\bfP_{S\cup F})}_{p,2}^p$ is small for every $k$-dimensional subspace $F$. 

\begin{Lemma}
Let $1 \leq p < \infty$ and $k\in\mathbb N$. There is an $s$-dimensional subspace $S$ where $s = O(k/\eps^{\max\{2, p\}})$ such that for every $k$-dimensional subspace $F$,
\[
    \norm{\bfA(\bfP_S-\bfP_{S\cup F})}_{p,2}^p \leq \eps^p \OPT
\]
where $\OPT = \min_{F\in\mathcal F_k}\norm{\bfA(\bfI-\bfP_F)}_{p,2}^p$.
\end{Lemma}
\begin{proof}
The proof largely follows \cite{SW2018} combined with our improved inequalities proved earlier. We reproduce a proof for the reader's convenience.

Lemma 6 in \cite{SW2018} shows that there is an $s$-dimensional subspace $S$ such that
\begin{equation}\label{eq:diff-bound}
    \norm{\bfA(\bfI-\bfP_S)}_{p,2}^p - \norm{\bfA(\bfI-\bfP_{S\cup F})}_{p,2}^p \leq \eps^{\max\{2,p\}} \OPT
\end{equation}
for every $k$-dimensional subspace $F\in\mathcal F_k$. We now use the fact that for any vector $\bfa\in\mathbb R^d$,
\[
    \norm{\bfa^\top(\bfP_S-\bfP_{S\cup F})}_2^2 = \norm{\bfa^\top(\bfI-\bfP_S)}_2^2 - \norm{\bfa^\top(\bfI-\bfP_{S\cup F})}_2^2
\]
by orthogonality and Lemma \ref{lem:p-diff} (with $\eps' = \eps^p$) to show that
\[
    \norm{\bfA(\bfP_S-\bfP_{S\cup F})}_{p,2}^p \leq \begin{cases}
        \eps^p \norm{\bfA(\bfI-\bfP_S)}_{p,2}^p + 2\eps^{p-2} (\norm{\bfA(\bfI-\bfP_S)}_{p,2}^p - \norm{\bfA(\bfI-\bfP_{S\cup F})}_{p,2}^p) & 1 \leq p < 2 \\
        \norm{\bfA(\bfI-\bfP_S)}_{p,2}^p - \norm{\bfA(\bfI-\bfP_{S\cup F})}_{p,2}^p & 2 \leq p < \infty
    \end{cases}
\]
by summing up the inequalities over vectors $\bfa_i$ for $i\in[n]$. By \eqref{eq:diff-bound}, we have that
\[
    \norm{\bfA(\bfP_S-\bfP_{S\cup F})}_{p,2}^p \leq 3\eps^p \OPT
\]
in any case. Rescaling $\eps$ by constant factors yields the statement of the theorem.
\end{proof}

Finally, we combine this bound with Lemma \ref{lem:proj-bound} to conclude Theorem \ref{thm:representative-subspace}.

\section{Efficient constructions of nearly linear representative subspaces}
\label{sec:fast-sw}

In this section, we sharpen a result of \cite{FKW2021} and show that representative subspaces of dimension $\tilde O(k)\eps^{-\Theta(p)}$ can be constructed in polynomial time. This implies that strong coresets of size $\tilde O(k^{\max\{1,p/2\}})\poly(\eps^{-1})$ \emph{with an added dimension} can be constructed in time $\tilde O(\nnz(\bfA)) + d\poly(k/\eps)$. Indeed, note that the work of \cite{HV2020, WY2023b} shows that strong coresets of size $\poly(k/\eps)$ can be constructed in input sparsity time via sensitivity sampling, so we may assume that $n$ and $d$ are at most $\poly(k/\eps)$. Thus, it suffices to design a polynomial time algorithm for constructing representative subspaces. For this, we use the observation of \cite{FKW2021} that we only need an efficient bicriteria algorithm that outputs a subspace which has objective value at most $(1+\eps)$ times the rank $k$ optimum. The work of \cite{FKW2021} uses an algorithm with bicriteria rank $\poly(k/\eps)$. We will show instead how to improve this to an algorithm with bicriteria rank $\tilde O(k/\eps)$.

We first start with a sketching result which sharpens \cite[Theorem 32]{CW2015a}.

\begin{Lemma}
\label{lem:right-sketch}
Let $\bfA\in\mathbb R^{n\times d}$. Let $\bfG$ be an $d\times r$ random Gaussian matrix for $r = \tilde O((k + \log(n/\delta))/\eps)$. Then,
\[
    \min_{\rank(\bfX)\leq k}\norm{\bfA\bfG\bfX - \bfA}_{p,2}^p \leq (1+\eps)\min_{\rank(F)\leq k}\norm{\bfA(\bfI-\bfP_F)}_{p,2}^p.
\]
\end{Lemma}
\begin{proof}
Let $\bfV^*$ be an orthonormal basis for the optimal rank $k$ projection $\bfP^*$, so that
\[
    \OPT = \norm{\bfA\bfV^*(\bfV^*)^\top - \bfA}_{p,2}^p = \min_{\bfX\in\mathbb R^{n\times k}}\norm{\bfX(\bfV^*)^\top- \bfA}_{p,2}^p.
\]
Note that the latter problem can be viewed as $n$ different $\ell_2$ linear regression problems. Then by sketching with $\bfG$ \cite{Woo2014}, for each $i\in[n]$, we have that with probability at least $1-\delta/n$ that $\hat\bfx = \bfe_i^\top\bfA\bfG ((\bfV^*)^\top\bfG)^-$ satisfies
\[
    \norm{\hat\bfx(\bfV^*)^\top - \bfe_i^\top\bfA}_2 \leq (1+\eps) \min_{\bfx\in\mathbb R^k}\norm{\bfx(\bfV^*)^\top - \bfe_i^\top\bfA}_2.
\]
By a union bound over the $n$ rows, this is simultaneously true for all $i\in[n]$ with probability at least $1-\delta$. Then by taking $p$-th powers and summing, we have
\[
    \norm{\bfA\bfG((\bfV^*)^\top\bfG)^-(\bfV^*)^\top - \bfA}_{p,2}^p \leq (1+O(\eps))\min_{\bfX\in\mathbb R^{n\times k}}\norm{\bfX(\bfV^*)^\top - \bfA}_{p,2}^p = (1+\eps)\OPT. \qedhere
\]
\end{proof}

The next lemma shows how Lemma \ref{lem:right-sketch} implies an efficient bicriteria algorithm of rank $\tilde O(k/\eps)$.

\begin{Lemma}
\label{lem:bicriteria}
There is a $\poly(nd)$ time algorithm which, given an input matrix $\bfA\in\mathbb R^{n\times d}$, outputs a rank $\tilde O(k/\eps)$ subspace $\hat F$ such that
\[
    \norm{\bfA(\bfI-\bfP_{\hat F})}_{p,2}^p \leq (1+\eps)\min_{\rank(F)\leq k}\norm{\bfA(\bfI-\bfP_F)}_{p,2}^p.
\]
\end{Lemma}
\begin{proof}
By Lemma \ref{lem:right-sketch}, we have that
\[
    \min_{\rank(\bfX)\leq k}\norm{\bfA\bfG\bfX - \bfA}_{p,2}^p \leq (1+\eps)\min_{\rank(F)\leq k}\norm{\bfA(\bfI-\bfP_F)}_{p,2}^p.
\]
Then, we can simply drop the rank $k$ constraint on $\bfX$ to solve for a matrix $\bfY$ of rank $\tilde O(k/\eps)$ such that $\norm{\bfA\bfY - \bfA}_{p,2}^p \leq (1+\eps)\OPT$. Note that $\bfY$ can be computed efficiently by, e.g., applying Dvoretzky's theorem to approximate the $(p,2)$-norm by a $(p,p)$-norm and then solving an instance of $\ell_p$ regression $d$ times. Finally, we can take $\hat F = \rowspan(\bfY)$ to produce a rank $\tilde O(k/\eps)$ subspace such that
\[
    \norm{\bfA(\bfI-\bfP_{\hat F})}_{p,2}^p \leq (1+\eps)\OPT.\qedhere
\]
\end{proof}

Finally the above bicriteria algorithm can be combined with the iteration algorithm idea of \cite{SW2018}, as done in \cite{FKW2021}, to prove the following theorem:

\begin{Theorem}
There is an algorithm which, given an input matrix $\bfA\in\mathbb R^{n\times d}$, runs in time $\poly(nd)$ and outputs a rank $\tilde O(k/\eps^{-\max\{2+p,2p\}})$ subspace $S$ such that
\[
    \norm{\bfA(\bfI-\bfP_F)}_{p,2}^p = (1\pm\eps)\norm{[\bfA\bfP_S(\bfI-\bfP_F),\bfb_S]}_{p,2}^p
\]
for every $F\in\mathcal F_k$, where $\bfb_S(i) = \norm{\bfa_i^\top(\bfI-\bfP_S)}_2$.
\end{Theorem}
\begin{proof}
To construct a linear size representative subspace in polynomial time, we will use the work of \cite{FKW2021}, which observes that $(1+\eps)$ approximation algorithms in the algorithm of \cite{SW2018} can be replaced by $(1+\eps)$ bicriteria approximation algorithms, which can in turn be implemented efficiently. The proof of this theorem now simply replaces the bicriteria algorithm of \cite{FKW2021} with our bicriteria algorithm of Lemma \ref{lem:bicriteria} and our sharpened analysis of \cite{SW2018} in Appendix \ref{sec:sharper-sohler-woodruff}. Note that to find a space $S$ such that
\[
    \norm{\bfA(\bfP_S - \bfP_{S\cup F})}_{p,2}^p \leq \eps \OPT
\]
the algorithm will add a rank $\tilde O(k/\eps)$ bicriteria space for $\eps^{-\max\{2/p,1\}}$ iterations, so the resulting subspace has rank at most $\tilde O(k)\eps^{-\max\{2/p + 1, 2\}}$. We need to apply this with $\eps$ replaced by $\eps^p$, so the result bound on the rank of the representative subspace is $\tilde O(k)\eps^{-\max\{2+p, 2p\}}$.
\end{proof}

\end{document}